\documentclass[twocolumn,showpacs,preprintnumbers,amsmath,amssymb]{revtex4}

\usepackage{pstricks}
\usepackage{graphicx}
\usepackage{dcolumn}
\usepackage{bm}

\begin{document}

\newcommand{\locsection}[1]{\setcounter{equation}{0}\section{#1}}
\renewcommand{\theequation}{\thesection.\arabic{equation}}

\def\F{{\bf F}}
\def\A{{\bf A}}
\def\J{{\bf J}}
\def\af{{\bf \alpha}}
\def\beqn{\begin{eqnarray}}
\def\eeqn{\end{eqnarray}}

\def\dspace{\baselineskip = .30in}
\def\beq{\begin{equation}}
\def\non{\nonumber}
\def\bwi{\begin{widetext}}
\def\ewi{\end{widetext}}
\def\pl{\partial}
\def\na{\nabla}
\def\al{\alpha}
\def\bt{\beta}
\def\Ga{\Gamma}
\def\ga{\gamma}
\def\de{\delta}
\def\De{\Delta}
\def\da{\dagger}
\def\ka{\kappa}
\def\si{\sigma}
\def\Si{\Sigma}
\def\te{\theta}
\def\La{\Lambda}
\def\lam{\lambda}
\def\Om{\Omega}
\def\om{\omega}
\def\ep{\epsilon}
\def\sq{\sqrt}
\def\sqg{\sqrt{G}}
\def\sp{\supset}
\def\sb{\subset}
\def\l{\left (}
\def\r{\right )}
\def\lq{\left [}
\def\rq{\right ]}
\def\fr{\frac}
\def\la{\label}
\def\hs{\hspace}
\def\vs{\vspace} 
\def\inf{\infty}
\def\ran{\rangle}
\def\lan{\langle}
\def\ov{\overline}
\def\tl{\tilde}
\def\tm{\times}
\def\lrar{\leftrightarrow}


\preprint{HD-THEP-03-56}

\title{Fermion Masses and Coupling Unification in $E_6$.
\\Life in the Desert}

\author{Berthold Stech}
\email{B.Stech@ThPhys.Uni-Heidelberg.DE}
\author{Zurab Tavartkiladze}%
 \email{Z.Tavartkiladze@ThPhys.Uni-Heidelberg.DE}
\affiliation{%
\it Institut f\"ur Theoretische Physik, Universit\"at Heidelberg,
Philosophenweg 16, \\
D-69120 Heidelberg, Germany
}%

\date{November 12, 2003}

\begin{abstract}

We present an $E_6$ Grand Unified model with a realistic pattern of
fermion masses. All standard model fermions are unified in three
fundamental $27$-plets (i.e. supersymmetry is not invoked), which 
involve in addition right handed neutrinos 
and three families of vector like heavy quarks and leptons. The lightest
of
those can lie in the low TeV range, being accessible to future collider
experiments. Due to the high symmetry, the masses and mixings of all
fermions are closely related. The new heavy fermions  play a crucial role
for the  quark and lepton mass matrices and the  bilarge neutrino
oscillations. In all channels generation mixing and ${\cal CP}$ violation
arise from a single antisymmetric matrix.
The $E_6$ breaking proceeds via an intermediate energy
region with $SU(3)_L\tm SU(3)_R\tm SU(3)_C$ gauge symmetry and a 
discrete left-right symmetry. This breaking pattern leads in a
straightforward
way to the  unification of the three gauge coupling constants at high
scales, providing for a long proton lifetime. The model also provides for
the unification of the top, bottom and tau Yukawa couplings and for new
interesting relations in flavor and generation space.

\end{abstract}

\maketitle

The exceptional group $E_6$ \cite{E6}, \cite{eric} is the preferred group
for Grand
Unification. All Standard Model (SM) fermions are in the lowest 
${\bf 27}$ representation. Its maximal subgroup 
$SU(3)\tm SU(3)\tm SU(3)$ can be viewed as an extension of the
Weinberg-Salam group 
$SU(2)_L\tm U(1)_Y\tm SU(3)_C\to SU(3)_L\tm SU(3)_R\tm SU(3)_C\equiv
G_{333}$. The fermions can be described by singlet and triplet
representations of the $SU(3)$ groups only. Using for all fermion fields
left handed two component Weyl spinor fields, the quantum number
assignments are (for
each generation) \cite{ft1}: 
\begin{eqnarray}
{\rm Quarks}:&~~~Q_L(x)=(3, 1, \bar 3)~,&\nonumber \\
{\rm Leptons}:&~~~L(x)=(\bar 3, 3, 1)~,&\nonumber \\
{\rm Antiquarks}:&~~~Q_R(x)=(1, \bar 3, 3)~.&
\la{assign}
\end{eqnarray}
The $78$ generators of $E_6$ consist of the three $SU(3)$ adjoint octet
generators and the generators $F(3, 3, 3)$ and 
$\bar F (\bar 3 , \bar 3 , \bar 3)$ of coset $E_6/G_{333}$.

The beautiful cyclic symmetry of $E_6$ is apparent from (\ref{assign}) 
and
from the fact that $F$ takes a quark field into a lepton field, a lepton
field into an antiquark field and an antiquark field into a quark
field. An additional argument favoring $E_6$ is its appearance 
through compactification of the ten dimensional $E_8\tm E_8$ heterotic
superstring theory on a Calabi-Yau manifold. The compactification process
can
lead either to four dimensional $E_6$ gauge symmetry (which is anomaly
free and left-right symmetric) or to some of $E_6$'s maximal 
subgroups \cite{CYE6}. 
Phenomenology of $E_6$ GUT attracted attention earlier 
\cite{E6}, \cite{eric},
\cite{antE6}, and its active studies has been continued until recently
\cite{newE6}.
Phenomenology and properties of $G_{333}$ triunification models are
also interesting \cite{SU3fromE6}.

According to (\ref{assign}) one has besides the SM fermions:
additional quark and antiquark
fields
with the same charges as the corresponding down quarks,
two $SU(2)_L$ doublet  leptons (containing additional
'active' neutrinos), and two SM singlets - 'right handed' neutrinos
for each generation.

$SO(10)$ and $E_6$ Grand Unified Theories in old times usually predicted
small neutrino mixings since in straightforward applications the large
symmetry  obtained from these groups connect the neutrino
mixings with the small mixings observed in the quark sector. After the
observation of large mixings in neutrino oscillations one had to return to
the smaller $SU(5)$ group (the minimal version of it does not involve
right
handed neutrinos) or needed several Higgses of the same representation, or 
special composite operators and fine tuning procedures.
In this paper we will show, however, that the consequent use of the 
fermion and scalar particle interactions and spectra of $E_6$ allows to 
construct a realistic GUT model.

We consider at first the Yukawa sector of $E_6$ with its
symmetric
and antisymmetric matrices in flavor and generation space. After defining
the model, we can calculate from it the mass spectrum of ordinary and new
fermions and their mixings in terms of a few parameters only. An
interesting feature is that the mass matrices of quarks and leptons are
strongly influenced by the flavor mixing of the SM particles with heavy
fermions as was suggested by Bjorken, Pakvasa and Tuan \cite{bpt}. 
Earlier suggestions for the mixing of the SM particles with new heavy
fermions  can be found in \cite{ros}.
Our work is done in the spirit of ref. \cite{bpt}. As in this reference, 
our scenario favors a relatively light mass scale for some of the new 
particles [$10$-plets of $SO(10)$]. The lightest can lie in the low TeV 
region or even below. A major difference to \cite{bpt} is the full use of 
the discrete left-right symmetry of $E_6$, valid at the intermediate 
symmetry $G_{333}$. It is broken solely by the Majorana property of 
very heavy neutral leptons (the right handed heavy neutrinos).
The use of an antisymmetric Higgs representation proposed many years ago
\cite{antE6} plays a decisive role. The corresponding antisymmetric 
matrix determines the generation mixing and the ${\cal CP}$ violation 
in all heavy and light channels (in the basis in which the up quark 
mass matrix is diagonal). The inclusion of all  neutral leptons
of $E_6$ allows to connect the mass matrix of the heavy neutrinos with 
the diagonal up quark mass matrix and the antisymmetric generation matrix. 
It leads to bimaximal mixings of the light neutrinos which then changes 
to a bilarge mixing pattern at the weak scale by renormalization effects.
All mass ratios and mixing angles
of light and heavy fermions are simply related to each other.

We then study the gauge coupling and top-bottom-tau unification in
$E_6$.
It is achieved by an unbroken $G_{333}$ subgroup as an intermediate
symmetry.
The discrete left-right ${\cal
D}_{LR}$ symmetry, which is unbroken at these intermediate energies, plays
also an important role. The breaking
scale $M_I$ of the intermediate symmetry is not a free parameter, but
uniquely fixed
by the standard model couplings: $M_I \simeq 1.5\cdot 10^{13}$~GeV.
$M_I$ also determines the scale
of light and heavy neutrinos in agreement with experiment. The unification
of the couplings occurs above $10^{16}$~GeV, in our specific model at
$2\cdot 10^{17}$~GeV and thus suppresses proton decay.
The renormalization of
mass ratios, various Yukawa matrices and the scaling of the neutrino mass 
matrix are studied in detail. 

Our model is non supersymmetric as the one in \cite{bpt}. The hierarchy 
problem persists but it is hoped that its eventual solution would not 
change the basic features of our approach.

\locsection{Particle Assignments in $E_6$ and the Yukawa Sector 
\label{sec:assig}}

Let us first consider the lowest particle generation
\begin{eqnarray}
&(Q_L)_i^a=
\left(
\begin{array}{c}
\hs{-0.3cm}u^a\\
\hs{-0.2cm}d^a\\
\hs{-0.2cm}D^a
\end{array}\right)~,~~
L^i_k=
\left(\begin{array}{ccc}
L^1_1 & E^{-} & e^{-} 
\\
E^{+} & L^2_2 & \nu
\\
e^{+} & \hat{\nu } & L^3_3
\end{array}\right)~,\nonumber \\
&~~(Q_R)^k_a=\l \hat{u}_a,~\hat{d}_a,~\hat{D}_a \r ~.
\label{partassign}
\end{eqnarray}
where $i, k, a=1, 2, 3$; $a$ is a color index.
In this description $SU(3)_L$ acts vertically and $SU(3)_R$
horizontally. The charges are obtained from the operator
\beq
Q=(I_3+\fr{1}{2}Y)_L+(I_3+\fr{1}{2}Y)_R~,
\la{oper}
\end{equation}
with $I_3$, $Y$ defined as usual. Before symmetry breaking 
equivalent forms of
(\ref{partassign}) can be obtained by applying left and right 
${\cal U}$-spin rotations.

The charge conjugation operator interchanges left with right handed
indices:
\begin{eqnarray}
&{\cal C}(Q_L)_i^a~{\cal C}^{-1}=(Q_R)^i_a~,~~~
{\cal C}L^i_k~{\cal C}^{-1}=L^k_i~,\nonumber \\
&{\cal C}(Q_R)^i_a~{\cal C}^{-1}=(Q_L)_i^a~.
\la{chconj}
\end{eqnarray}
It leaves the commutation relations for the $E_6$ generators unchanged and
is often called 
${\cal D}_{LR}$ parity.
The new lepton fields $L^1_1$, $L^2_2$, $L^3_3$ are identical with their
own antiparticle fields. 
Nevertheless, if two of these fields, say $L^1_1$ and
$L^2_2$, are connected by a single mass term a four component Dirac field
can be formed. The two fields then behave like a (vector
like) particle-antiparticle pair.
The parity and ${\cal CP}$ operations change the left handed two component
fields into right handed ones:
\begin{eqnarray}
&{\cal P}(Q_L)_i^a(t, {\rm x}){\cal P}^{-1}=
\si_2(Q_R)^{i*}_a(t, -{\rm x})~,\nonumber \\
&{\cal P}L^i_k(t, {\rm x}){\cal P}^{-1}=
\si_2L^{k*}_i(t, -{\rm x})~,\nonumber \\
&{\cal P}(Q_R)^i_a(t, {\rm x}){\cal P}^{-1}=
\si_2(Q_L)^{a*}_i(t, -{\rm x})~,\nonumber \\
&{\cal CP}(Q_L)_i^a(t, {\rm x}){\cal CP}^{-1}=
\si_2(Q_L)^{a*}_i(t, -{\rm x})~,\nonumber \\
&{\cal CP}L^i_k(t, {\rm x}){\cal CP}^{-1}=
\si_2L^{i*}_k(t, -{\rm x})~,\nonumber \\
&{\cal CP}(Q_R)^i_a(t, {\rm x}){\cal CP}^{-1}= 
\si_2(Q_R)^{i*}_a(t, -{\rm x})~.
\la{PCPoparat}
\end{eqnarray}

By including the generation quantum number $\al $($=1, 2, 3$) all basic
fermions are now classified by the left handed Weyl fields $\Psi^{\al }_r$
with the $E_6$ flavor index $r$ running from $1$ to $27$.

The product of two ${\bf 27}$'s of $E_6$ decomposes into a symmetric 
$\ov{\bf 27}$, an antisymmetric $\ov{\bf 351}_A$ and a symmetric 
$\ov{\bf 351}_S$ representation
\beq
{\bf 27}\tm {\bf 27}=\ov{\bf 27}+\ov{\bf 351}_A+\ov{\bf 351}_S~.
\la{dec27prod}
\end{equation}
Consequently, the Yukawa interactions in the $E_6$ Lagrangian 
contein in general the three Higgs fields
\beq
H=H(27)~,~~H_A=H(351_A)~,~~H_S=H(351_S)~.
\la{higgses}
\end{equation}
Each of
the three Higgs fields couple to the fermions together with a $3\tm 3$
matrix $G$ acting on
the generation space
\beqn
&G_{\al \bt }=[G(27)]_{\al \bt }~,~~~
A_{\al \bt }=[G(351_A)]_{\al \bt }~,\nonumber \\
&S_{\al \bt }=[G(351_S)]_{\al \bt }~.
\la{Gmatr}
\eeqn
The $E_6$ invariant Yukawa interaction reads
\beqn
&{\cal L}_{Y}=\l (\Psi^{\al }_r)^T{\rm i}\si_2\Psi^{\bt }_s\r
\lq G_{\al \bt }H_{rs}+A_{\al \bt }(H_A)_{rs}+ \right. \nonumber \\
&\left.  S_{\al \bt }(H_S)_{rs}\rq +{\rm h.c.}
\la{Yukawa}
\eeqn
$G$ and $S$ are symmetric matrices in generation space, while $A$ is an
antisymmetric matrix. 
${\cal L}_{Y}$ is invariant with respect to ${\cal C}$, the 
right$\lrar $left operation. 
In case of real vacuum expectation values (VEVs) of the Higgs fields,
the part of ${\cal L}_{Y}$ obtained from the real
part of these matrices is formally even under the ${\cal CP}$ and 
${\cal P}$ operation, while the term arising from their imaginary parts is 
formally odd under ${\cal CP}$ and ${\cal P}$.

The decomposition of the Higgs fields with respect to the $G_{333}$
subgroup reads
\beq
H=(\bar 3, 3, 1)+(1, \bar 3, 3)+(3, 1, \bar 3)~,
\la{decH333}
\end{equation}
\beqn
&H_A=(\bar 3, 3, 1)+(1, \bar 3, 3)+(3, 1, \bar 3)+\nonumber  \\
&(\bar 3, \bar 6, 1)+(1, \bar 3, \bar 6 )+(\bar 6, 1, \bar 3)+
(6, 3, 1)+(1, 6, 3)+\nonumber \\
&(3, 1, 6)+(3, 8, \bar 3 )+(\bar 3, 3, 8)+(8, \bar 3, 3)~,
\la{decHA333}
\eeqn
\beqn
&H_S=(\bar 3, 3, 1)+(1, \bar 3, 3)+(3, 1, \bar 3)+\non \\
&(6, \bar 6, 1)+(1, 6, \bar 6)+(\bar 6, 1, 6)+\non \\
&(3, 8, \bar 3)+(\bar 3, 3, 8)+(8, \bar 3, 3)~.
\la{decHS333}
\eeqn
The color singlet parts who's neutral members can develop VEVs are
\beqn
&H\to (\bar 3, 3, 1)~,\non \\
&H_A\to (\bar 3, 3, 1)+(\bar 3, \bar 6, 1)+(6, 3, 1)~,\non \\
&H_S\to (\bar 3, 3, 1)+(6, \bar 6, 1)~.
\la{VEVcomps}
\eeqn
We note that the parts containing a sextet or antisextet representation
can only couple to leptons.

\locsection{The Model\label{sec:model}}

The vacuum expectation values of the three Higgs fields determine the
particle spectrum. To be in accord with the SM the masses of the new
particles of $E_6$ have to get  heavy (at least 
of order TeV). Thus, Higgs
components which are $SU(2)_L$ singlets can have large VEVs. The members
of $SU(2)_L$ doublets, on the other hand, should be of the order of the
weak scale, while the VEVs of $SU(2)_L$ triplets are expected to vanish.

In order to define our model to be predictive and to have very few unknown
parameters, we need some specific assumptions concerning the
three Higgs fields, about the generation matrices $G$, $A$, $S$ and
the symmetry breaking pattern. We do not consider Higgs field components
which carry
color. They are supposed to aquire masses of the order of the GUT scale
from appropriate Higgs potentials.

We allow VEVs for all color singlet and neutral
components of $H$
\beq
\lan H_1^1\ran =e^1_1~,~~~\lan H^i_k\ran =e^i_k~,~~~{\rm for}~i, k=2, 3~.
\la{HVEVs}
\end{equation} 
However, by a biunitary left and right ${\cal U}$-spin transformation in
flavor space (i.e. on the $SU(3)_L$ and $SU(3)_R$ indices $2$ and $3$) we
can choose a proper basis for which
\beq
e^2_3=e^3_2=0.
\la{VEVbasis}
\end{equation}
Our first assumption concerns the VEVs of
$H_A$ and $H_S$. $\lan H_A\ran $ can mix the standard model particles with
the new heavy $D$ and $L$ states [the $10$-plet of $SO(10)$]:
$d\lrar D$, $e\lrar E$, $\nu =L^2_3\lrar L^2_2$. This is achieved by
components of $H_A$ which involve left and right ${\cal U}$-spin $1/2$
indices. For the $(\bar 3, 3, 1)$ sector of $H_A$ we take, therefore,
\beq
\lan (H_A)^i_k\ran =f^i_k~,~~~~~
~~i,k=2, 3~.
\la{AVEVs}
\end{equation}
For the $(\bar 3, \bar 6, 1)$ sector of $H_A$ one has correspondingly
\beq
\lan H_A^{i\{1, k\}}\ran =f^{i\{1, k\}}~,~~~~~ ~~i,k=2, 3~,
\la{Abar6VEVs}
\end{equation}
and for the sector $(6, 3, 1)$
\beq
\lan H_{A\{1, k\}i}\ran =f_{\{1, k\}i}~,~~~~~ ~~i,k=2, 3~.
\la{A6VEVs}
\end{equation}
In our numerical treatment we will restrict the VEVs in (\ref{Abar6VEVs}),
(\ref{A6VEVs}) to those with $i=3$, $k=2, 3$ and $i=2$, $k=3$ which should
be the dominant ones.
With respect to ${\cal U}$-spin, $f^{3\{1, 3\}}$ is the analogue of
$f^3_2$. While $f^3_2$ mixes $d$ with $D$, $f^{3\{1, 3\}}$ mixes 
$e^{-}$ with $E^{-}$ and $\nu $ with $L^2_2$.

The VEVs of the symmetric Higgs field $H_S$ can provide large Majorana
masses for the heavy leptons $L^3_2$ and $L_3^3$. They arise from the
$H_S(6, \bar 6, 1)$ sector. Here we have to take the $SU(2)_L$ singlets
and left and right handed ${\cal U}$-spin triplets
\beq
\lan (H_S)_{\{3, 3 \}}^{\{2, 2 \}}\ran =F^{\{2, 2 \}}~,~~~~~
\lan (H_S)_{\{3, 3 \}}^{\{3, 3 \}}\ran =F^{\{3, 3 \}}~.
\la{ASVEVs}
\end{equation}
All other components of $\lan H_A\ran $ and $\lan H_S\ran $ are taken 
to be zero or negligible in our calculations.

Of particular interest is the question of the breaking of the left-right
symmetry of $E_6$. $F^{\{22\}}$ as obtained from $\lan H_S\ran $ breaks
this symmetry strongly. It could be the dominant manifestation of 
${\cal D}_{LR}$ symmetry breaking. $\lan H\ran $ and $\lan H_A\ran $ on
the other hand need not break this symmetry significantly. A strict left
right symmetry in this sector would imply the relations
\beq
f^i_k= -f^k_i~,~~~f^{i\{1, k\}}= f_{\{1, k\}i}~,~~~~~i,k=2,3~.
\la{lrVEVs}
\end{equation}

The signs follow by taking the $H_A$ part of the Yukawa interaction
to be even under ${\cal D}_{LR}$ when $H_A$ is replaced by
$\lan H_A\ran $.
As a consequence of (\ref{lrVEVs}) the f's are of the order of the
weak scale even though some are
standard model singlets and thus only protected by the discrete
${\cal D}_{LR}$ symmetry itself. 
If this is indeed the case, it implies new particles in the few TeV
region as we will see.

The next assumption concerns the generation matrices $G$, $A$ and $S$. The
symmetric matrix $G_{\al \bt }$ can be diagonalized by an orthogonal
transformation, which leaves the symmetry properties of $A_{\al \bt }$
and $S_{\al \bt }$ unchanged. By choosing this basis, the up quark mass
matrix is diagonal because, according to the above assumed properties of
$\lan H_A\ran $ and $\lan H_S\ran $, only $\lan H\ran $ contributes to it
\beq
\l m_U\r_{\al \bt}=G_{\al \bt }e^1_1=
g_{\al }\de_{\al \bt }e^1_1~.
\la{upM}
\end{equation}
As a consequence, the quark mixing angles and the ${\cal CP}$ violating
phase must entirely come from the inclusion of the Higgs $H_A$ with its
antisymmetric generation matrix $A_{\al \bt }$ as proposed in ref.
\cite{antE6}. Thus, $A_{\al \bt }$ has to contain imaginary parts which
can not be rotated away using quark phase redefinitions. This leads us to
assume that the matrix $A$ is - in our phase convention - purely
imaginary, i.e. a hermitian matrix. The normalized matrix contains then
only two parameters, in fact only one when utilizing a discrete generation
exchange symmetry for $A$ as shown later.

We suggest, that the generation matrices
$G$, $A$ and $S$ are not independent of each other. 
In particular, the coupling matrix $S$ for the heaviest leptons
should have an intimate relation with the generation matrices of the
charged
fermions \cite{sipar}.
$S$ may then be expanded in terms of $G$ and $A$. Speculatively we assume:
The generation mixing matrix $S$ is a combination of the
bilinear product $G^2$ and the commutator $[G, A]$.
The generation mixing in this sector is then due to the same matrix $A$
which causes the mixing of the charged fermions. 
As it turns out this structure for $S$ is crucial
for bilarge neutrino mixings. 
In fact, it leads to bimaximal mixing which is then changed to bilarge
mixing by renormalization group effects.

The last assumption concerns the breaking pattern of
$E_6$, which is presumably the origin of the breakings seen in the Yukawa
sector.
We suppose the following symmetry breaking chain:
\beqn
&E_6\stackrel{M_{\rm GUT}}{\longrightarrow}  
SU(3)_L\tm SU(3)_R\tm SU(3)_C\tm {\cal D}_{LR}    
\stackrel{M_I}{\longrightarrow}\non \\
&  SU(2)_L\tm U(1)_Y\tm SU(3)_C~.
\la{chain}  
\eeqn
Here $M_{\rm GUT}$ is the GUT scale and ${\cal D}_{LR}$ denotes the discrete
left$\lrar $right symmetry operation.
As we will show below, the breaking chain (\ref{chain}) 
leads in a straightforward way to the unification of the gauge coupling
constants.
The first breaking step to the intermediate symmetry can be caused by a
scalar $650$-plet which contains
two $G_{333}$ singlets. One of them ${\cal S}_{+}$  is even under 
${\cal D}_{LR}$ (${\cal S}_{+}\to {\cal S}_{+}$), while the second one 
${\cal S}_{-}$ is odd (${\cal S}_{-}\to -{\cal S}_{-}$). 
It then follows from the symmetries at $M_I$ and above $M_I$ that
${\cal S}_{+}$ has the non zero VEV 
 and
$\lan {\cal S}_{-}\ran =0$. This insures that in the
$(M_I,~M_{\rm GUT})$ interval $L\lrar R$ symmetry is precise and the
equality 
of the coupling constants $g_L$ and $g_R$ is protected also at the quantum
level. 

We take two Higgs $SU(2)_L$ doublets of $H$, namely $H^{1,2}_1$ and
$H^{1,2}_2$, to be relatively light. The remaining Higgs masses of the
color neutral components of $H$ and $H_A$ are taken to be of order $M_I$
or higher. The only exception is the $SU(2)_L$ doublet $(H_A)^{1,2}_2$
which can be much lighter than $M_I$ because of the left right symmetry
($f^2_2 \approx 0$) in the $H_A$ sector mentioned above. But it must be
heavier than $\approx 500$~TeV not to induce flavor changing processes
above presently known limits.  All
Higgses not mentioned are assumed to have masses at the order of the GUT
scale.

Before starting our investigation, let us state the quark and lepton
masses at the scale $\mu =M_Z$ \cite{jamin}, \cite{parida} 
\beqn
&m_u=(1.8\pm 0.4)~{\rm MeV}~,~~~~
m_d=(3.3\pm 0.7)~{\rm MeV}~,\non \\
&m_s=(62\pm 12)~{\rm MeV}~,~~~~
m_c=(0.64\pm 0.04)~{\rm GeV}~,\non \\
&m_b=(2.89\pm 0.03)~{\rm GeV}~,~~~~ 
m_t=(173\pm 5)~{\rm GeV}~,\non \\
&m_e=0.487~{\rm MeV}~,~~~~
m_{\mu }=102.8~{\rm MeV}~,\non \\
&m_{\tau }=1.747~{\rm GeV}~,
\la{expQLmassMZ}
\eeqn  
as obtained from the analysis of experimental data. The general
hierarchical structure of the SM masses and of the CKM matrix elements
will be used in the following. Some of the masses, in particular $m_t$,
$m_b$ and $m_{\tau }$, are taken as input parameters.

\locsection{The Quark Mass Matrix\label{sec:quark}}

Because of the hierarchical structure of the quark masses and mixing
angles it is convenient to express them in terms of powers of a small
dimensionless parameter. We introduce the parameter $\si $ \cite{sipar}
with the value
\beq
\si =0.058~,
\la{si}
\end{equation}
for which
\beq
|V_{cb}|\simeq \fr{\si }{\sq{2}}~~{\rm and}~~
\left | \fr{m_u}{m_c}\right |\simeq 
\left | \fr{m_c}{m_t}\right |\simeq |V_{ub}|\simeq \si^2
\la{hiers}
\end{equation}
holds within experimental uncertainties. One also has 
$\fr{m_s}{m_b}|V_{us}|\approx \si^2$~.

According to our assumption (\ref{HVEVs}), (\ref{VEVbasis}) the up-quark
mass matrix is
\beqn
&\l m_U\r_{\al \bt}\equiv G_{\al \bt}\lan H^1_1\ran =
g_{\al }\de_{\al \bt }e^1_1=\non \\
&{\rm Diag }\l \fr{m_u}{m_t},~\pm \fr{m_c}{m_t}, ~1 \r m_t~.
\la{upq}
\eeqn
At the scale $\mu\simeq M_Z$ we can write
\beq
m_U\simeq {\rm Diag}\l \si^4,~\pm \si^2,~1 \r 173~{\rm GeV}~.
\la{upqMz}
\end{equation}
The signs of the mass parameters are in general of no relevance
because of the freedom to change phases. But since
we keep $G$ and $A$ to be hermitian matrices, the Jarlskog determinant
obtained from the commutator of mass matrices depends on the sign chosen
in (\ref{upq}) giving two solutions for the area of the unitarity
triangle.

Because of the existence of the $D$-quarks, the down quark (big) mass
matrix is a $6\tm 6$ matrix, which contains the antisymmetric
generation matrix $A$
\begin{equation}
\begin{array}{cc}
 & {\begin{array}{cc}
 \hspace{-0.1cm}\hat{d}\hspace{2.2cm}& \hspace{1.5cm}\hat{D}
\end{array}}\\ \vspace{2mm}
M_{d, D}=
\begin{array}{c}
 d\\D  
\end{array}\!\!\!\!\!\! &{\left(\begin{array}{ccc}
\hs{-0.1cm}e^2_2G+f_2^2A~,&\hs{-0.1cm} f_3^2A
\\
\hs{-0.2cm}f_2^3A~, &\hs{-0.1cm}e^3_3G+f_3^3A
\end{array} \right)\! }~.
\end{array}  
\label{MdD}
\end{equation} 
Here $e^2_2$, $f_2^2$ and $f_3^2$ are mass scales of order of the weak
scale,
while at least $e^3_3$ should describe a heavy mass scale. In accord
with our model assumptions we have
$f^3_2, f^3_3\ll e^3_3$ which allows to integrate out the $D, \hat{D}$
states and to write down the see-saw formula

\beq
m_D\simeq e^2_2G+f_2^2A-\fr{f_2^3f_3^2}{e^3_3}\l AG^{-1}A\r ~.
\la{Md}
\end{equation}
The $D$-quark mass matrix is simply proportional to the up quark mass
matrix:
\beq
M_D\simeq e^3_3G=\fr{e^3_3}{e^1_1}m_U~.
\la{MD}
\end{equation}
Although eq. (\ref{MdD}) should only be valid at the unification scale and
has to be carefully scaled down for a determination of $m_D$ at $\mu
\simeq M_Z$, we will use eq. (\ref{Md}) at $M_Z$ for a first orientation.

The first entry in (\ref{Md}) is responsible for the mass of the bottom
quark, while the second term must provide for the
small mixing angles and the large ${\cal CP}$ violating phase. 
The third term  gives a correction to the symmetric part of
the mass matrix which is important for the strange quark mass.
We expect, therefore,
\beqn
&e_2^2\simeq m_b~,~~~f_2^2|A_{23}|\simeq |V_{cb}|m_b~,\non \\
&f_2^2|A_{12}|\simeq m_s|V_{us}|~,~~{\rm and}~~
f_2^2|A_{13}|\simeq |V_{ub}|m_b~.
\la{fit1}
\eeqn
{}From (\ref{hiers}) one then gets for $f^2_2$ and the generation 
matrix $A$
\begin{equation}
\begin{array}{cc}
 & {\begin{array}{cc}
 \hspace{-0.1cm}\hspace{2.2cm}& \hspace{-0.5cm}
\end{array}}\\ \vspace{2mm}
f_2^2\simeq \fr{\si m_b}{\sq{2}\lam_A}~,~~~~~~A\simeq \hs{-0.2cm}
\begin{array}{c}
 \\  
\end{array}\!\!\!\!\!\! &{\left(\begin{array}{ccc}
0~,&{\rm i}\si ~, &-{\rm i}\si
\\
-{\rm i}\si ~, &0~, &\fr{{\rm i}}{\sq{2}}
\\
{\rm i}\si ~,&-\fr{{\rm i}}{\sq{2}}~, &0
\end{array} \right)\lam_A\sq{2}\! }~,
\end{array}  
\label{f1A}
\end{equation}
We introduced a scaling factor $\sq{2}\lam_A$ (as discussed in section
\ref{sec:unif}) such that $f_2^2=\lan H^2_2\ran $ and the scale 
dependent matrix $A$ is
normalized according to ${\rm Tr}(A^2)\simeq 2\lam_A^2$. 
We remark that the
antisymmetric matrix $A$ taken here is also 
antisymmetric with respect to the (discrete) interchange of the second
generation with the third one.  
We know of course, that the matrix $A$ can have its strictly antisymmetric
form only above $M_I$, the breaking point of the left-right
symmetry. Thus, in our renormalization group treatment we take the matrix
$A$ as given in eq. (\ref{f1A}) to be strictly valid at 
$\mu =M_{\rm GUT}$, even though we anticipated its form at a
low scale. As we will discuss in the appendix, by going down from $\mu
=M_{\rm GUT}$ to $M_I$, the matrix $A$ 'splits' into a matrix $A^Q$ for
the quarks and a matrix $A^L$ for the leptons. By going further down to 
$M_Z$, $A^Q$ as well as $A^L$ each splits into three matrices relevant for
the sectors indicated by the superscripts:
\beqn
&A^Q\to \l A^{d\hat{d}}~,~A^{d\hat{D}}~,~A^{D\hat{d}}\r ~~~
{\rm and}~~~\non \\
&A^L\to \l A^{e^{-}e^{+}}~,~A^{e^{-}E^{+}}~,~A^{E^{-}e^{+}}\r ~. 
\la{AQLsplit}
\eeqn
These matrices are no more strictly antisymmetric.
Obviously, also the matrix $G$ splits into more matrices. Between
$M_{\rm GUT}$ and $M_I$ we have $G\to \l G_Q,~G_L\r$ for the quarks and
leptons. Below $M_I$, one gets
\beqn
&G_Q\to \l G^{u\hat{u}}~,~G^{d\hat{d}}~,~G^{D\hat{D}}\r 
~~~{\rm and}~~~\non \\ 
&G_L \to \l G^{e^+e^{-}}~,~G^{L\bar L}~,~G^{\nu \hat{\nu }}\r ~.
\la{GQLsplit}
\eeqn

We calculated these matrices at $\mu =M_Z$ in a model specified in section
\ref{sec:unif}, which has a unification scale of $2\cdot 10^{17}$~GeV. The 
matrices $A$ are
exhibited in (\ref{matrAdec1})-(\ref{23ALMZ}). 
The matrix $e_2^2G$ in (\ref{Md}) becomes $e_2^2G^{d\hat{d}}$.
It is obtained from $m_U=e_1^1G^{u\hat{u}}$ at $M_Z$ [eq. (\ref{upqMz})]
scaled up to the GUT scale, where $G^{u\hat{u}}=G^{d\hat{d}}$ holds and
then scaled
down to $M_Z$. In our approximation, it is still a diagonal matrix.
In the same way, one obtains the matrix replacing
$G^{-1}$ in the $D\hat{D}$ channel of eq. (\ref{Md}). It is denoted by 
$(G^{D\hat{D}})^{-1}$. The matrices  $G^{D\hat{D}}$ and $G^{d\hat{d}}$ at
$\mu =M_Z$ are given in (\ref{MDDspace}), (\ref{LDatMZ}).

With these changes the mass matrix for the down quarks becomes now
$$
m_D(M_Z)=e_2^2G^{d\hat{d}}+f_2^2A^{d\hat{d}}-
\fr{f_2^3f_3^2}{e_3^3}
A^{d\hat{D}}(G^{D\hat{D}})^{-1}A^{D\hat{d}}~.
%
$$
\beq
e_2^2(G^{d\hat{d}})_{33}\simeq m_b^0~,~~~~~
f_2^2(A^{d\hat{d}})_{23}\simeq \fr{{\rm i}\si m_b^0}{\sq{2}}~.
\la{renMd}
\end{equation}
$m_b^0$ will slightly differ from the mass of the bottom quark because of 
the mixing occuring in $m_D$.

After having found the renormalization group effects on the matrices $G$
and $A$,    
the only  parameter for calculating the $d$-quark masses and
the CKM  matrix is $f_2^3f_3^2/(e^3_3(G^{D\hat{D}})_{33})$. We use this
parameter for a fit of
the
Cabibbo angle $|V_{us}|$. Because of our expectation of an approximate
left right symmetry [see (\ref{lrVEVs})] we look for a negative value of
this parameter and find
\beq
\fr{f^2_3f^3_2}{e^3_3(G^{D\hat{D}})_{33}}
\simeq -4.75\cdot 10^{-5}~{\rm GeV}~.
\la{fixf2f3e33}
\end{equation}
Upon diagonalization of
the down quark mass matrix (\ref{renMd}), with the negative sign taken in
(\ref{upqMz}), one obtains
\beqn
&m_d(M_Z)\simeq 2.66~{\rm MeV}~,~~~~
m_s(M_Z)\simeq 49.7~{\rm MeV}~,\non \\
&m_b(M_Z)\simeq 2.89~{\rm GeV}~,\non \\
&|V_{us}|\simeq 0.217~,~~|V_{cb}|\simeq 0.045~,~~
|V_{ub}|\simeq 0.0034~,~~~~~~~
\la{downCKM}
\eeqn
and for the angles of the unitarity triangle
\beq
\al \simeq 84^{o}~,~~~~~\bt \simeq 20^{o}~,~~~~~\ga \simeq 76^{o}~.
\la{unitr}
\end{equation}
To obtain the correct value for $m_b(M_Z)$ we took
for the (3,3) element of $e_2^2G^{d\hat{d}}$, 
$m_b^0=2.859$~GeV.
A similar good fit is obtained if in (\ref{upqMz}) the positive sign is
chosen. The
number given in (\ref{fixf2f3e33}) then changes to 
$-3.26\cdot 10^{-5}$~GeV 
and the
angles of the unitarity triangle become
    \beq
\al \simeq 95^{o}~,~~~~~\bt \simeq 21^{o}~,~~~~~\ga \simeq 64^{o}~.
\la{unitr2}
\end{equation}
In the following we will use the negative sign in (\ref{upqMz}).

The results (\ref{downCKM})-(\ref{unitr2}) are in good agreement with
present experimental data. The mass of the strange quark is a bit
low but still within the bounds of (\ref{expQLmassMZ}).
We also see, that Weinberg's suggestion
\cite{wein}
\beq
|V_{us}|\approx \sq{\fr{m_d}{m_s}}~,
\la{pred1}
\end{equation}
is valid. It follows from the smallness of the $(1, 1)$ entry $e_2^2\si^4$
in (\ref{renMd}) due to the small first generation up quark mass.
We further note, that the term in $m_D$, which arises from the mixing
with the
heavy $D$-quarks, reduced the angle $\ga $ from the originally obtained
value  $\simeq 90^{o}$ \cite{antE6} to a lower value.

Besides (\ref{fixf2f3e33}) there is no restriction on the value of
$e^3_3$ except that $f^3_2/e^3_3$ has to be sufficiently small to justify
the see-saw formula and thereby the near unitarity of the CKM mixing
matrix.
However, as mentioned in sect. 2, the VEVs
$\lan H\ran$ and $\lan H_A\ran$ may approximately respect the left-right
symmetry of
$E_6$ and of the intermediate symmetry in contrast to 
the large VEV of $H_S$. This idea is
supported by the small value
found for $f^2_2$ in
(\ref{f1A}). It would be zero for a strict
left right symmetry in this channel and is indeed small
($f^2_2 \simeq 0.093$~GeV) compared to
the weak interaction scale. One can then expect, that the product
$-f^2_3 f^3_2$ is not of order $M_Z M_I$ but not much higher than 
$(M_Z)^2 $. This
gives us a rough estimate for $e^3_3$ and thus for the masses of the
$ D$ quarks.
\beq
e^3_3(G^{D\hat{D}})_{33} 
\approx \fr{2.1\cdot 10^4}{\rm GeV}M_Z^2~~{\rm or}~~
\approx \fr{3.07\cdot 10^4}{\rm GeV}M_Z^2~.
\la{LRf2f3}
\end{equation}
From these relations, which are of course sensitive to the value taken
for the weak scale input, we expect 
$e_3^3(G^{D\hat{D}})_{33}$ to be of order $(10^7-10^8)$~GeV.
Taking $e_3^3(G^{D\hat{D}})_{33}=4\cdot 10^7$~GeV as an example 
(and scaling effects into account), one obtains
\beqn
&M_{D1} \simeq 557~{\rm GeV}~,~~~~ 
M_{D2} \simeq 129~{\rm TeV}~,\non \\
&M_{D3} \simeq 4\cdot 10^4~{\rm TeV}~.
\eeqn
A more detailed discussion of the heavy fermions and their masses  
will be presented in section \ref{sec:desert} and in the appendix A1.

\locsection{The Charged Lepton Mass Matrix\label{sec:chl}}

The charged lepton mass matrix has the same structure as the down quark
mass matrix. 
By going from quarks to leptons $E_6$ Clebsch-Gordon coefficients have to
be taken into account.
Quarks and leptons couple to $H(\bar 3, 3, 1)$ according to the
combination
\beq
q_i\hat{q}^k+
\fr{1}{2}\varepsilon_{ii'i''}
\varepsilon^{kk'k''}L^{i'}_{k'}L^{i''}_{k''}~.
\la{glebsh}
\end{equation} 
The $(\bar 3, 3, 1)$ sector of the Higgs field $H_A$ couples only to
quarks, the sectors $(\bar 3, \bar 6, 1)$ and $(6, 3, 1)$ only to leptons.
Thus, the relevant $6\tm 6$ matrix at the GUT scale is
\begin{widetext}
\begin{equation}
\begin{array}{cc}
 & {\begin{array}{cc}
 \hspace{-0.1cm}e^{+}\hspace{4.5cm}& \hspace{3.5cm}E^{+}
\end{array}}\\ \vspace{2mm}
M_{e, E}=
\begin{array}{c}
 e^{-}\\E^{-}  
\end{array}\!\!\!\!\!\! &{\left(\begin{array}{ccc}
\hs{-0.1cm}-e^2_2G\hs{-1mm}-\hs{-1mm}(f^{2\{1,3\}}\hs{-1mm}-
\hs{-1mm}f_{\{1,3\}2})A~,&
\hs{-0.1cm} f^{3\{1,3\}}A
\\
\hs{-0.2cm}-f_{\{1,3\}3}A~, &\hs{-0.1cm}-
e^3_3G\hs{-1mm}+\hs{-1mm}(f^{3\{1,2\}}\hs{-1mm}-\hs{-1mm}f_{\{1,2\}3})A
\end{array} \right)\! }~.
\end{array}  
\label{MeE}
\end{equation}
\end{widetext}
Using the same arguments as for the down quark mass matrix the $f$'s in
the diagonal elements are small compared to the main terms.
After integrating out the
$E$-type states, the mass matrix for the charged leptons of the SM is
generated and has at $\mu \simeq M_Z$ the form
\beqn
&m_E\simeq - e_2^2G^{e^{-}e^+}-
(f^{2\{13\}}-f_{\{13\}2})A^{e^{-}e^{+}} \non \\
&-\fr{f_{\{1,3\}3}f^{3\{1,3\}}}{e^3_3} 
A^{e^{-}E^{+}}(G^{L\bar L})^{-1}A^{E^{-}e^{+}} ~.
\la{Me}
\eeqn
The first term is constructed like $e_2^2G^{d\hat{d}}$, but for leptons
and given in the appendix A2.
The contribution of VEVs in the second term should be as small as the
corresponding term $f_2^2$ in the
quark mass matrix.
Diagonalizing (\ref{Me}) one gets with
\beqn
&f^{2\{13\}}-f_{\{13\}2}\simeq 0.042~{\rm GeV}~~~{\rm and }\non \\
&\fr{f_{\{1,3\}3}f^{3\{1,3\}}}{e^3_3(G^{L\bar L})_{33}}
\simeq 12.6\cdot 10^{-5}~{\rm GeV}~, 
\la{inpLep}
\eeqn
the charged lepton masses
\beq
m_e=0.488~{\rm MeV}~,~~~~m_{\mu }=102.8~{\rm MeV}~,~~~~
m_{\tau }=1.748~{\rm GeV}~.
\la{mefit}
\end{equation}
{}For obtaining the correct value of the tau lepton mass  we took
$m_{\tau }^0$ [the (3,3) element  of
$e_2^2G^{e^{-}e^+}$] to be $1.689$~GeV.
The contributions from the first term in (\ref{Me}) for
the light generations are proportional to $\si^4$ and
$\si^2$, respectively, and thus negligeably small.
The muon mass receives its essential
contribution from the third term in (\ref{Me}). i.e. from the mixing with
the heavy leptons. The contribution from the second and third terms
to the electron mass are comparable.
There is some ${\cal CP}$-violation due to the second term in (\ref{Me}).
The corresponding unitarity triangle, for charged leptons, has the angles:
$~\al \simeq 43^{o}~,~\bt \simeq 62^{o}~,~\ga \simeq 75^{o}~$.
After diagonalization of the charged lepton matrix, this ${\cal CP}$
violation will affect the neutrino mixings.
The charged lepton mixing angles turn out to be small
$|V_{e\mu }|\simeq 0.034$, $|V_{e\tau }|\simeq 0.003$, 
$|V_{\mu \tau }|\simeq 0.068$.
Therefore, the large neutrino mixings are not due to the mixings in the
charged lepton sector but 
should come from the neutral lepton sector
where  large Majorana masses appear.
In the next section it will be shown that this is indeed the case.
Comparing now (\ref{fixf2f3e33}) with (\ref{inpLep}) and taking into
account (\ref{RATcases}) we get
$\fr{f_{\{1,3\}3}f^{3\{1,3\}}}{-f^2_3f^3_2}\simeq (1.6)^2$.
Considering the analogy of $f^2_3$ with $f_{\{1,3\}3}$ and $f^3_2$ with
$f^{3\{1,3\}}$ this appears to be a reasonable value.
In sections \ref{sec:desert}, \ref{sec:unif} we will use
$|f_{\{1,3\}3}|\simeq |f^{3\{1,3\}}|$,
$|f_3^2|\simeq |f_2^3|$, i.e. appropriate left-right symmetry 
in these channels.

\locsection{The Neutral Lepton Mass Matrix\label{sec:nu}}

The fundamental fermion representation of $E_6$ contains five neutral
two-component fields. Thus, for three generations, the mass matrix for
these neutral leptons is a $15\tm 15$ matrix. According to the assumption
stated in section \ref{sec:model}, it is given by
\begin{widetext}
\begin{equation}
\begin{array}{ccccc}
 & {\begin{array}{ccccc}
\hspace{0cm} L^2_3\hspace{1cm} & \hspace{1cm}L_2^3 \hspace{1cm} 
& \hspace{1.3cm}L_3^3\hspace{1cm} &\hspace{1.1cm}L^1_1\hspace{1cm} &
\hspace{1cm}L^2_2\hspace{0.8cm}  
\end{array}}\\ \vspace{2mm}
M_L\hs{-0.1cm}=\hs{-0.2cm}
\begin{array}{c}
L^2_3 \\ L^3_2 \\ L^3_3 \\ L_1^1 \\ L^2_2
 \end{array}\!\!\!\!\! \hs{-0.1cm}&{\left(\begin{array}{ccccc}
0 &\hs{-0.1cm}-e^1_1G &\hs{-0.1cm}0 &
\hs{-0.2cm} -f^{3\{1,3\}}A &\hs{-0.2cm}0
\\  
\hs{-0.2cm}-e^1_1G   &
\hs{-0.1cm}F^{\{ 2, 2\}}S  &\hs{-0.1cm}0 &
\hs{-0.2cm}-f_{\{1,3\}3}A &\hs{-0.2cm}0
 \\
0 &\hs{-0.1cm}0 &\hs{-0.1cm}F^{\{ 3, 3\}}S &
\hs{-0.2cm}e^2_2G\hs{-0.7mm}+\hs{-0.7mm}h_2^2A  &
\hs{-0.2cm}e^1_1G
\\
\hs{-0.2cm}-f^{3\{1,3\}}A^T &\hs{-0.1cm}-f_{\{1,3\}3}A^T
& 
\hs{-0.1cm}e^2_2G\hs{-0.7mm}+\hs{-0.7mm}h_2^2A^T 
& \hs{-0.2cm}0  &\hs{-0.2cm}e^3_3G\hs{-0.7mm}+\hs{-0.7mm}h_3^3A 
\\
0 & \hs{-0.1cm}0 &\hs{-0.1cm}e^1_1G & 
\hs{-0.2cm}e^3_3G\hs{-0.7mm}+\hs{-0.7mm}h_3^3A^T
& \hs{-0.2cm}0
\hs{-0.1cm}\end{array}\hs{-0.1cm}\right)}~,
\end{array}  \!\!  
\label{ML} 
\end{equation}
\end{widetext}
where 
\beq
h_2^2=f^{2\{1,3\}}+f_{\{1,3\}2}~,~~~~h_3^3=f^{3\{1,2\}}+f_{\{1,2\}3}~,
\la{defh}
\end{equation}
and
$L^2_3$ stands for the standard light neutrino fields.
All ingredients in this matrix arising from the Higgs fields $H$ and $H_A$
are defined in the previous sections. We notice, however, that in the 
$L_1^1L_2^2$ block the contribution of the $f$s is additive. 
The new elements are the
ones containing the symmetric
generation matix $S$. They give rise to genuine Majorana mass terms
and are of particular significance in the diagonalization process. 
The strength of the $H_S$ Higgs contribution to $M_L$ is governed
by the constants $F^{\{ 2, 2\}}$ and
$F^{\{ 3, 3\}}$ carrying right handed ${\cal U}$-spin  quantum
numbers. $F^{\{ 3, 3\}}$ essentially fixes the Majorana mass for the heavy
$L^3_3$ leptons which are expected to be of the order of the
$SU(3)_L\tm SU(3)_R$ breaking scale.
The constant $F^{\{ 2, 2\}}$ of similar strengths breaks
the left-right symmetry  and thus is responsible for the dominant breaking 
of this symmetry.

We can reduce the matrix $M_L$ to a $9\tm 9$ matrix by knowing that
$e_3^3$ is
much larger than the other elements in the same row and column, in
particular, if $f^{3\{1,3\}}$ and $f^{3\{1,2\}}$ are indeed of the order
of the weak
scale. This allows to integrate out the $L_1^1$, $L_2^2$ states. With
the abbreviations
\beq
f=f^{3\{1,3\}}e_1^1/e_3^3~,~~~
\bar f=f_{\{1,3\}3}e_1^1/e_3^3~,
\la{abr}
\end{equation}
one finds
\begin{equation}
\begin{array}{ccc}
 & {\begin{array}{ccc}
\hspace{-0.6cm} ~~L^2_3\hspace{0.7cm} & \hspace{0.7cm}L_2^3 \hspace{1.1cm} 
& \hspace{0.6cm}L^3_3\hspace{2cm}  
\end{array}}\\ \vspace{1mm}
{M_L}'=
\begin{array}{c}
L^2_3 \\ L^3_2 \\ L^3_3 
 \end{array}\!\!\!\!\! &{\left(\begin{array}{ccccc}
0 &-e^1_1G &fA
\\  
-e^1_1G   &F^{\{2, 2 \}}S  &
\bar fA
 \\
fA^T  &\bar fA^T &
F^{\{3, 3 \}}S
\end{array}\right)}~,
\end{array}  \!\!  ~~~~~
\label{ML1}
\end{equation}
and
\beq
M_{L_1^1L_2^2}\approx e_3^3G~.
\la{ML11L22}
\end{equation}
We neglected  in (\ref{ML1}) a correction to the (3-3) block, namely
$-\fr{e^1_1}{e^3_3}e^2_2G$.
It is small compared to the large eigenvalues  of
$F^{\{3,3\}}S$. 

{}For values of $f^{3\{1,3\}}$ of the order of the weak scale and 
$F^{\{(2,2)\}}, F^{\{(3,3)\}}$ near $M_I\approx 10^{13}$~GeV, we
can again apply the see-saw mechanism and  finally arrive
at the
$3\tm 3$ Majorana matrix for the light neutrinos 
\beq
m_{\nu }\simeq -\fr{(e^1_1)^2}{F^{\{2,2\}}}GS^{-1}G-
\fr{f^2}{F^{\{3, 3 \}}}A^TS^{-1}A~,
\la{Mnu}
\end{equation}
and for the mass matrices of the heavy Majorana neutrinos
\beq
M_{L_2^3}\simeq F^{\{2,2\}}S~,~~~~~
M_{L_3^3}\simeq F^{\{3,3\}}S~.
\la{masMajnu}
\end{equation}
Only the first term in (\ref{Mnu}) need to be
considered, since the
remaining one can safely be neglected. 
Therefore, the neutrino mass matrix (\ref{Mnu}) is
mainly due to the decoupling of $L_2^3=\hat{\nu }$ states. It scales
with the masses of these heavy lepton states. 

We expect 
[see sect. 2] $S$ to be related to the two other
generation matrices $G$ and $A$.
The main term for $S$ should be $G^2$, which leads to a diagonal non degenerate
mass matrix [see (\ref{Mnu})]. We then add a term linear in $A$,
the commutator $[G, A]$, with a tiny coefficient.
It implies that also in this sector generation mixing is 
solely due to the antisymmetric matrix $A$. We take for $S$, 
divided by the overall coupling strength $\lam_S$ to the Higgs field
$H_S$, the real and bilinear construct
\beq
S/\lam_S=G^2+{\rm i}x\si^3 [G, A]~,
\la{bilin}
\end{equation}
whith the single parameter $x$. The $G^2$ term with its dominant 
element $\simeq 1$ for the $3^{\rm rd}$ generation serves for generation
hierarchy and for the normalization of $S/\lam_S$ (for which the $\si^3$
term in (\ref{bilin}) can be neglected).
With no renormalization effects included, the matrix $S$, as defined in
(\ref{bilin}) reads
\begin{equation}
\begin{array}{cc}
 & {\begin{array}{cc}
 \hspace{-0.1cm}\hspace{2.2cm}& \hspace{-0.5cm}
\end{array}}\\ \vspace{2mm}
S/\lam_S\simeq \hs{-0.2cm}
\begin{array}{c}
 \\  
\end{array}\!\!\!\!\!\! &{\left(\begin{array}{ccc}
\si^8~,&-\si^6x ~, &-\si^4x
\\
-\si^6x ~, &\si^4~, &\fr{\si^3}{\sq{2}}x
\\
-\si^4x ~,&\fr{\si^3}{\sq{2}}x~, &1
\end{array} \right)\! }~.
\end{array}  
\label{Sform}
\end{equation}                
In each element of (\ref{Sform}) 
only the leading powers of $\si $ are shown.

By inverting the matrix $S$ defined in (\ref{bilin}) and using
(\ref{Mnu}), one finds for $m_{\nu }$ 
\begin{widetext}
\beq
\begin{array}{cc}
 & {\begin{array}{cc}
 \hspace{-0.1cm}\hspace{2.2cm}& \hspace{-0.5cm}
\end{array}}\\ \vspace{2mm}
m_{\nu }\simeq \hs{-0.1cm}
\begin{array}{c}
 \\  
\end{array}\!\!\!\!\!\! &{\left(\begin{array}{ccc}
-1~,&(1-\fr{1}{\sq{2}}\si x)x ~, & -(1-\fr{1}{\sq{2}}\si x)x
\\
(1-\fr{1}{\sq{2}}\si x)x  ~, &x^2-1~, 
&(x-\fr{1}{\sq{2}}\si )x
\\
-(1-\fr{1}{\sq{2}}\si x)x~, 
&(x-\fr{1}{\sq{2}}\si )x , & x^2-1
\end{array} \right)
\fr{(e_1^1\lam_{\tau })^2}{
(1-2x^2+\sq{2}x^3\si )\lam_SF^{\{2,2\}} }\! }~.
\end{array}  
\label{Mnubare}
\end{equation}
\end{widetext}
{}For the simplicity of representation (\ref{Mnubare}) contains only the
zeroth and first powers in $\si $. Taking the full
expression makes numerically little difference. 
The interesting feature of $m_{\nu }$ is the fact that it produces for any
value of $x\stackrel{>}{_\sim }1.5$  automatically an 
almost perfect bimaximal neutrino mixing pattern (!) with a normal (not 
inverted) neutrino spectrum. 
By changing $x$, solely the ratio of mass square differences
\beq 
R=\fr{m_2^2-m_1^2}{m_3^2-m_2^2}~,
\la{nuratio}
\end{equation}
changes ($m_i$ denotes the three $\nu $
eigenstates mass ordered according to $m_1<m_2<m_3$). 
The experimentally observed ratio $R\approx 0.03 $ is obtained for
$x\simeq 3.5$.

However, for a proper calculation of the neutrino mass matrix at $\mu
=M_I$ and $\mu =M_Z$, renormalization effects have to be taken into
account. This is particularly necessary because of the large generation
splitting of the heavy neutrino states $L_2^3=\hat{\nu }$ caused by the 
$G^2$ term in the matrix $S$. We have to integrate out these states in
steps and to redefine $m_{\nu }$ in each step.
We start by using (\ref{bilin}) at the scale $M_I$ 
with $G=G_L$ and $A=A_L$ and proceed according to the rules given in 
appendix A4. 
It turns out that renormalization effects strongly influence the 
neutrino mass matrix and thus also the mixing pattern. The bimaximal 
mixing 
is changed to a bilarge mixing. The calculation is again performed for
the gauge and Yukawa unification at $2\cdot 10^{17}$~GeV described in 
section \ref{sec:unif}. 
As in the examples given in \cite{ratz} we find that the renormalization
coefficients strongly reduce the mixing angle $\te_{12}$ observed in 
solar neutrino experiments while the angle $\te_{23}$ observed in 
atmospheric neutrino experiments is less affected. The renormalization 
coefficients also increase the value of the ratio $R$.

A good description of the known neutrino data is obtained by changing the 
value of our parameter $x=3.5$ to
\beq
x\simeq 2.8~.
\la{newrenx}
\end{equation}
With this value we obtain $R\simeq 0.055$. Larger values of $x$ reduce 
$R$. However, this would lead to a too strong reduction of the solar 
neutrino oscillation probability.

With  $x\simeq 2.8$ one obtains for the mass matrix of the light 
neutrinos at $\mu =M_Z$ 

\begin{equation}
\begin{array}{cc}
 & {\begin{array}{cc}
 \hspace{-0.1cm}\hspace{2.2cm}& \hspace{-0.5cm}
\end{array}}\\ \vspace{2mm}
m_{\nu }= \hs{-0.1cm}-\hs{-0.3cm}
\begin{array}{c}
 \\  
\end{array}\!\!\!\!\!\! &{\left(\begin{array}{ccc}
-0.135~,&0.67 ~, & \hs{-3mm}-0.62
\\
0.67~,  &3.75~, &4.61
\\
-0.62~,&4.61 ~, &2.81
\end{array} \right)\fr{M_0}{10}\! }~,
\end{array}  
\label{Mnumatr}
\end{equation}
with 
\beq
M_0=
\fr{(\lam_{\tau }(M_I)e_1^1(M_I))^2}
{\lam_SF^{\{2,2\}}}\simeq 
\fr{(102.17~{\rm GeV})^2}{M_{\hat{\nu}_3\hat{\nu}_3}}~.
\la{defM0}
\end{equation}
Here $\lam_{\tau }$ is the coupling of the third generation lepton to
$H_2^{1,2}$ and $e_1^1$ is the VEV of the Higgs field 
$H_1^1$ as used before.

We obtain the mass squared difference observed in atmospheric neutrino 
experiments 
$\De m^2_{\rm atm}\simeq 2.5\cdot 10^{-3}~{\rm eV}^2$
when setting $M_{\hat{\nu }_3\hat{\nu }_3}\simeq 1.6\cdot 10^{14}$~GeV.
It is very satisfying that this scale is of the same order of magnitude as
expected from the value of $M_I$, the breaking point of the left-right
symmetry \cite{ft2}.
With this value of $M_{\hat{\nu }_3\hat{\nu }_3}$ the neutrino mass
eigenvalues turn out to be
\beqn
&m_1=0.0023~{\rm eV}~,~~m_2=0.0120~{\rm eV}~,~~ m_3=0.0516~{\rm eV}~,\non \\
&m_2^2-m_1^2\simeq 1.4\cdot 10^{-4}~({\rm eV})^2~,\non \\
&m_3^2-m_2^2\simeq 2.5\cdot 10^{-3}~({\rm eV})^2~.
\la{lnumasses}
\eeqn

To obtain the neutrino mixing matrix, one has to go to a basis in which
the charged lepton mass matrix is diagonal. Diagonalizing (\ref{Me}) and
denoting by $\nu_{\al }$ the
weak eigenstates ($\al =e, \mu , \tau $), we find
from (\ref{Mnumatr})
\beq
\nu_{\al }=U^{\nu }_{\al i}\nu_i ~,
\la{weakmass}
\end{equation}
\begin{equation}
\begin{array}{cc}
 & {\begin{array}{cc}
 \hspace{-0.1cm}\hspace{2.2cm}& \hspace{-0.5cm}
\end{array}}\\ \vspace{2mm}
U^{\nu }_{\al i}\simeq \hs{-0.2cm}
\begin{array}{c}
 \\   
\end{array}\!\!\!\!\!\! &{\left(\begin{array}{ccc}
-0.2+0.86{\rm i}~,&-0.46-0.08{\rm i} ~, &0.026
\\
-0.29-0.015{\rm i} ~, &0.044+0.55{\rm i}~, &0.78
\\
0.37-0.017{\rm i} ~,&-0.036-0.69{\rm i}~, &0.62
\end{array} \right)_{\al i}\! }~.
\end{array}  
\label{Unu}%
\end{equation}
Here we took a special phase choice for the neutrino flavor eigenstates
such that
the $3^{\rm rd}$ column has only real and 
positive elements.

Our results for the three mixing angles relevant in neutrino 
oscillation experiments as obtained from eq. (\ref{Unu}) are
\beq
\sin^2 \te_{13}\simeq 6.8\cdot 10^{-4}~,~~~
\sin^2 \te_{12}\simeq 0.22~,~~~
\sin^2 \te_{23}\simeq 0.61~.
\la{tetaagnl}
\end{equation}
These values and the ratio $R\simeq 0.055$ of mass squared differences are
quite close to the result \cite{valle} of
SuperKamiokande \cite{SK1},  \cite{SK2}, SNO \cite{sno} and KamLAND
\cite{KamL}, the
CHOOZ limit \cite{chooz} and the observations of the 
disappearance of solar neutrinos \cite{HSG}.

We can also get from (\ref{Unu}) the neutrino unitarity triangle, defined 
in analogy with the quark unitarity triangle. It turns out to be:
\beq
\al_{\nu }\simeq 74^o~,~~~~
\bt_{\nu }\simeq 6^o~,~~~~
\ga_{\nu }\simeq 100^o~.
\la{lepUniTr}
\end{equation}
The phases of the elements of the first row of $U^{\nu }$ are 
'Majorana phases' relevant for neutrinoless double $\bt $-decay 
experiments.
With the convention used in (\ref{Unu}) we get
\beq
\de_{11}\simeq 103~,~~~~
\de_{12}\simeq -170^o~,~~~~
\de_{13}= 0~.
\la{btphs}
\end{equation}
{}For the the quantity $\lan m_{\nu }\ran_{ee}$ which
determines the decay rate 
we find
\beqn
&\lan m_{\nu }\ran_{ee} 
=\left |m_1(U^{\nu }_{e1})^2+m_2(U^{\nu }_{e2})^2
+m_3(U^{\nu }_{e3})^2 \right |\simeq \non \\ 
&9.5\cdot 10^{-4}~{\rm eV}~.
\la{Mnuee}
\eeqn

The matrix $U^{\nu }$, in particular its deviation from bimaximal
mixing, depends via the renormalization parameters 
to some extent on the 
way unification is obtained. But a bilarge mixing with near maximal mixing 
in the $\mu -\tau $ sector will always result from the basic assumptions 
of our $E_6$ model outlined in section \ref{sec:model}.

\vs{0.3cm}

\locsection{The Desert is Blooming\label{sec:desert}}

In our model the masses of the heavy down quarks $D$ and the
corresponding leptons $L$, which form a $10$-plets of $SO(10)$,  
have a generation splitting similar to the up quarks. The absolute values
of
these masses can not be given. However, if $\lan H_A\ran $ still respects
to some extent
the left right symmetry of $E_6$ as discussed above, the lightest $D$ and
$L$ states lie in the TeV region. In section \ref{sec:unif}, we 
present a numerical
solution of the problem of the gauge  and Yukawa coupling unification for 
$M_{\rm GUT}=2\cdot 10^{17}$~GeV and 
$e_3^3(G^{D\hat{D}})_{33}=3.98\cdot 10^{7}$~GeV. This 
solution also fixes
the so far undetermined VEVs of $H$ and $H_A$. With
the values quoted there [eq. (\ref{VEVsmod})], we can now directly
diagonalize the $6\tm 6$
matrices (\ref{MdD}) and (\ref{MeE}) which determine the mixing of the
SM particles with the $D$ and $L$ states. This mixing, although important
for the mass matrices, does not seriously violate the unitarity relations
for the SM particles. For example, the sum of the
squares of the second row of the CKM matrix
differs from one only by $1.2\cdot 10^{-4}$. In the charged lepton
sector, the corresponding deviation amounts to $3.7\cdot 10^{-4}$.

We can list now the mass values of the new particles 
by using again $e_3^3(G^{D\hat{D}})_{33}= 4\cdot 10^7$~GeV and setting
$\lam_SF^{\{2, 2 \}}=\lam_SF^{\{3, 3 \}}=1.6\cdot 10^{14}$~GeV in accord
with the neutrino results:
\beqn
&M_{D_1}\simeq 557~{\rm GeV}~,~~~~
M_{D_2}\simeq 129~{\rm TeV}~,\non \\
&M_{D_3}\simeq 4\cdot 10^{4}~{\rm TeV}~,\non \\
&M_{L_1}\simeq 355~{\rm GeV}~,~~~~M_{L_2}\simeq 103~{\rm TeV}~,\non \\
&M_{L_3}\simeq 3.83\cdot 10^{4}~{\rm TeV}~,\non \\
&M(L^3_2)_1\simeq 5.8\cdot 10^{5}~{\rm GeV}~,~
M(L^3_2)_2\simeq 9.6\cdot 10^{8}~{\rm GeV}~,\non \\
&M(L^3_2)_3\simeq 1.6\cdot 10^{14}~{\rm GeV}~,\non \\
&M(L^3_3)_1\simeq 5.8\cdot 10^{5}~{\rm GeV}~,~
M(L^3_3)_2\simeq 9.6\cdot 10^{8}~{\rm GeV}~,\non \\
&M(L^3_3)_3\simeq 1.6\cdot 10^{14}~{\rm GeV}~. 
\la{MasBloom}
\eeqn 
In the evaluation we took the most important renormalization effects into
account (see section \ref{sec:unif} and the appendix).
As we see, the desert is populated between the mass scales
$M_Z$ and $M_{\rm GUT}$.
The mass ratios for different generations of the standard model singlet
neutrinos are even more drastic than the corresponding ratios for the $D$
quarks and the $SU(2)_L$ doublet heavy leptons.

Our specific unification model allows to calculate numerous properties of
the old and new particles in particular those related to their decay
properties. We will present here a few examples only.

{}From the $6\tm 6$ mass matrix  (\ref{MdD}), for the quarks, one can
calculate 
the coupling matrices in generation space for the couplings of the light
and heavy mass eigenstates to the appropriate light Higgs field components 
\beqn
&d^T\hs{0.1cm}{\rm i}\si_2{\cal C}^{d\hat{d}}\hs{0.1cm}\hat{d}
\hs{0.1cm}H_2^2~,~~~~~
d^T\hs{0.1cm}{\rm i}\si_2{\cal C}^{d\hat{D}}\hs{0.1cm}\hat{D}
\hs{0.1cm}H_2^2~,\non \\
&D^T\hs{0.1cm}{\rm i}\si_2{\cal C}^{D\hat{d}}\hs{0.1cm}
\hat{d}\hs{0.1cm}H_2^2~.
\la{DqH}
\eeqn 
We find, without using the remaining freedom of changing phases, 
\bwi
\beqn
&\begin{array}{cc}
 & {\begin{array}{cc}
 \hspace{-0.1cm}\hspace{2.2cm}& \hspace{-0.5cm}
\end{array}}\\ \vspace{2mm}
{\cal C}^{d\hat{d}}\simeq \hs{-0.4cm}
\begin{array}{c}
 \\  

\end{array}\!\!\!\!\!\! &{\left(\begin{array}{ccc}
(-0.18+1.6\hs{0.5mm}{\rm i})\cdot 10^{-4}~,&
(-0.25+7.3\hs{0.5mm}{\rm i})\cdot 10^{-4} ~, &
-(3.2+0.9\hs{0.5mm}{\rm i})\cdot 10^{-5}
\\
(7.4+0.28\hs{0.5mm}{\rm i})\cdot 10^{-4} ~, &(3.3+1.3\hs{0.5mm}{\rm i})\cdot
10^{-3}~,
& 
(-0.42+1.5\hs{0.5mm}{\rm i})\cdot 10^{-4}
\\
-(9.9+0.4\hs{0.5mm}{\rm i})\cdot 10^{-3} ~,&-0.043+0.014\hs{0.5mm}{\rm i} ~,
&0.069+{\rm i}
\end{array} \right)\! }~,
\end{array}  \non \\
&
\begin{array}{cc}
 & {\begin{array}{cc}
 \hspace{-0.1cm}\hspace{2.2cm}& \hspace{-0.5cm}
\end{array}}\\ \vspace{2mm}
{\cal C}^{d\hat{D}}\simeq \hs{-0.4cm}
\begin{array}{c}
 \\  

\end{array}\!\!\!\!\!\! &{\left(\begin{array}{ccc}
0~,&-4.4\cdot 10^{-7} ~, &0
\\
-(0.3+3.4\hs{0.5mm}{\rm i})\cdot 10^{-2} ~, &0~, &
4.1{\rm i}\cdot 10^{-6}
\\
-0.77-8.9\hs{0.5mm}{\rm i} ~,&0.39 ~, &0
\end{array} \right)\cdot 10^{-3}\! }~,
\end{array}  \non \\
&
\begin{array}{cc}
 & {\begin{array}{cc}
 \hspace{-0.1cm}\hspace{2.2cm}& \hspace{-0.5cm}
\end{array}}\\ \vspace{2mm}
{\cal C}^{D\hat{d}}\simeq \hs{-0.3cm}
\begin{array}{c}
 \\  
\end{array}\!\!\!\!\!\! &{\left(\begin{array}{ccc}
0~,&(0.33-3.8\hs{0.5mm}{\rm i})\cdot 10^{-2} ~,
&0.84-9.7\hs{0.5mm}{\rm i}
\\
-4.8\cdot 10^{-7} ~, &0~, &0.43
\\
0 ~,&-4.6\cdot 10^{-6}~, &0
\end{array} \right)\cdot 10^{-3}\! }~.
\end{array}  
\label{Cmatr}
\eeqn 
\ewi
Of course, similar results can be derived for the lepton couplings to
the Higgs fields.

{}For the weak interaction process $D\to u\hs{0.1cm}W_L$ one can introduce
the matrix ${\bf V}^{uD}$ as an extension of the CKM matrix
\beq
\ov{u}\hs{0.1cm}\ga^{\mu }(1-\ga_5){\bf V}^{uD}\hs{0.1cm}D
\hs{0.1cm}(W_L)_{\mu }^{+}~.
\la{uDW}
\end{equation}
{}From (\ref{MdD}) one gets
\begin{equation}
\begin{array}{cc} 
 & {\begin{array}{cc}
 \hspace{-0.1cm}\hspace{2.2cm}& \hspace{-0.5cm}
\end{array}}\\ \vspace{2mm}
{\bf V}^{uD}\simeq \hs{-0.3cm}
\begin{array}{c}
 \\  
\end{array}\!\!\!\!\!\! &{\left(\begin{array}{ccc}
-3.5\cdot 10^{-4}~,&-0.04 ~, &-1.1\cdot 10^{-4}
\\
-0.95+11\hs{0.5mm}{\rm i} ~, &0~, &1.3\cdot 10^{-3}
\\
0.81-9.4\hs{0.5mm}{\rm i} ~,&0.42 ~, &0
\end{array} \right)\cdot 10^{-3}\! }~.
\end{array}  
\la{VuD}
\end{equation}
There are also right handed current interactions of the standard model 
particles with the heavy $SU(2)_R$ vector bosons $W_R^{\pm }$
\beq
\ov{u}\hs{0.1cm}\ga^{\mu }(1+\ga_5){\bf V}^{\hat{u}\hat{d}}\hs{0.1cm}d
\hs{0.1cm}(W_R)_{\mu }^{+}~,
\la{udWR}
\end{equation}
where ${\bf V}^{\hat{u}\hat{d}}$ is slightly different but has the same
structure as the CKM matrix.

Of particular interest for the decay properties of the mass
eigenstates of $\hat{\nu }$
neutrinos are the Dirac masses connecting the flavor eigenstates of the
light neutrinos (in a basis in which the charged lepton matrix is
diagonal) with the heavy neutrinos. Using (\ref{Me}),
$e_1^1G_L(M_I)$ from (\ref{YLMI}), (\ref{1sole11}), (\ref{ka12a}) and
diagonalizing $S(M_I)$ we obtain
\bwi
\beq
\begin{array}{cc}
 & {\begin{array}{ccc}
 \hat{\nu}_1\hspace{0.5cm}&\hspace{2.5cm}\hat{\nu}_2\hspace{0.3cm} &
\hspace{2.2cm}\hat{\nu}_3\hspace{1cm}
\end{array}}\\ \vspace{2mm}
m^{\rm Dirac}\simeq 
\begin{array}{c}
\nu_1 \\  
\nu_2 \\
\nu_3
\end{array}\!\!\!\!\!\! &{\left(\begin{array}{ccc}
(1.7-10\hs{0.5mm}{\rm i})\cdot 10^{-4}~,&(-1.2-11\hs{0.5mm}{\rm i})\cdot
10^{-3} ~,
&-0.41-0.0099\hs{0.5mm}{\rm i}
\\
(-4.6+0.3\hs{0.5mm}{\rm i})\cdot 10^{-3}~, &0.033+0.32\hs{0.5mm}{\rm i}~,
&3.8+7.7\hs{0.5mm}{\rm i}
\\
(3.3+3.9\hs{0.5mm}{\rm i})\cdot 10^{-3} ~,&-0.034+0.041\hs{0.5mm}{\rm i}~,
&-93.7+83.6\hs{0.5mm}{\rm i}
\end{array} \right) {\rm GeV}\! }~.
\end{array}  
\label{CnuN}
\end{equation}
\ewi

\locsection{Unification of Couplings\label{sec:unif}}

\subsection{Gauge coupling unification 
with intermediate \\
$SU(3)_L\tm SU(3)_R\tm SU(3)_C$ symmetry\label{sec:unifG}}
 
As it is known, the SM does not lead to the unification of the gauge
coupling constants. In our scenario, there are
the additional Dirac fermions $D$ and $L$ below the GUT scale $M_{\rm GUT}$. 
However, these do not alter the
unification picture of the standard model significantly. We still need to
introduce an intermediate breaking scale $M_I$. 

A large group like $E_6$ with high dimensional representations should
first be
broken by a step which lowers the symmetry considerably. 
It is natural to break $E_6$ to the maximal subgroup 
$SU(3)_L\tm SU(3)_R\tm SU(3)_C$. As we will see, this has the advantage
that the corresponding intermediate scale is not an arbitrary parameter
but fixed. The breaking at the GUT scale can be achieved in the scalar
sector by a Higgs
$H(650)$, which contains two $G_{333}$ singlets $(1, 1, 1)$,
${\cal S}_{+}$ and ${\cal S}_{-}$. ${\cal S}_{+}$ is even under 
${\cal D}_{LR}$ and thus keeps the left-right symmetry, while 
${\cal S}_{-}$ is odd. We have to take $\lan {\cal S}_{-}\ran =0$ and
$\lan {\cal S}_{+}\ran $ to be different from zero for the
breaking. It keeps
$g_L=g_R$ [$g_L=g_{SU(3)_L}$, $g_R=g_{SU(3)_R}$] for 
$\mu \stackrel{>}{_-}M_I$. The reason is, that at 
the intermediate scale $M_I$ the $SU(2)_L$ gauge coupling $g_2(\mu )$ and
the hypercharge coupling $g_1(\mu )$ have to respect the 
$SU(3)_L\tm SU(3)_R$ symmetry. Since the $U(1)_Y$ hypercharge is a
combination of $Y_L$, $I_{3R}$ and $Y_R$, according to (\ref{oper}), the
intermediate symmetry automatically requires the matching
\beqn
&g_2(M_I)=g_1(M_I)=g_L(M_I)=g_R(M_I)\non \\
&{\rm and}~~g_L(\mu )=g_R(\mu )~~{\rm for}~~\mu \stackrel{>}{_-}M_I.
\la{matching}
\eeqn
The relation $g_L(\mu )=g_R(\mu )$ for $\mu \stackrel{>}{_-}M_I$ holds
even at the quantum level since it is protected by 
${\cal D}_{LR}$ parity. As a consequence, $M_I$ is fixed by the
meeting point of $g_2$ and $g_1$. From thereon the two curves continue as
a single one up to $M_{\rm GUT}$ where $g_L=g_R$ unifies with 
$g_C=g_{SU(3)_C}$. For this to happen the states $H_A(6, 3, 1)$ and
$H_A(\bar 3, \bar 6, 1)$ will play a central role as
we will see shortly.

The details are as follows: Below $M_I$, the field content consists of the
fermionic generations of the standard model together with two light Higgs
doublets and the three Dirac particles
$D(D,~\hat{D})$, $L^0(L^1_1, L_2^2)$, $L^{-}(L^1_2, L_1^2)$. The two Higgs
doublets are $H_1^{1,2}$, $H_2^{1,2}$ with
$\lan H_1^1\ran^2+\lan H_2^2\ran^2 =v_0^2
\stackrel{<}{_-}v^2=(174~{\rm GeV})^2$. $v_0$ will be smaller than $v$ in
case an additional Higgs meson with standard model quantum numbers has a
non zero VEV.

The corresponding $b$-factors for the evolution of the couplings are
\beq
\l b_1, b_2, b_3\r = 
\l \fr{21}{5}, -3, -7\r ~,
\la{bSM}
\end{equation}   
as obtained from the standard model fermions and two Higgs doublets.
The additional $b$-factors for the $D$'s and $L$'s for each generation are
\beq 
\l b_1, b_2, b_3 \r^D=\l \fr{4}{15}, 0, \fr{2}{3}\r~,~~
\l b_1, b_2, b_3 \r^L=\l \fr{2}{5}, \fr{2}{3}, 0 \r ~. 
\la{bAD} 
\end{equation} 
Apart from these additional states, there are more scalar doublets
$H_d^m$ ($m=1, 2, 3, 4$), which are involved in the construction of
the fermion sector. 
One of them comes from $H(\bar 3, 3, 1)$ and the other three from 
$H_A(\bar 3, 3, 1)$. $H$ and $H_A$ also contain two isosinglet fields
$H_{+}^n$ ($n=1, 2$) carrying the same $U(1)_Y$ charge as $e^{+}$.
If some of the corresponding states
with masses $\mu (H_d^m)$ and $\mu (H_{+}^m)$  lie below $M_I$, each
of them will contribute to the $b$-factors according to 
\beq
\l b_1, b_2, b_3 \r^{H_d}=\l \fr{1}{10}, \fr{1}{6}, 0\r ~,~
\l b_1, b_2, b_3 \r^{H_{+}}=\l \fr{1}{5}, 0, 0\r .
\la{addoubl}
\end{equation}
Four more Higgs components which are SM singlets could also  be relatively 
light, but they do not
contribute to the running of the gauge couplings.
Thus, the solution of the renormalization group equation (at one loop
level) for the gauge couplings at $M_I$ reads
\beqn
&\al_a^{-1}(M_I)=\al_a^{-1}(M_Z)-\fr{b_a}{2\pi }\ln \fr{M_I}{M_Z}\non \\
&-\fr{b_a^{H_d}}{2\pi }\sum_{m}\ln \fr{M_I}{\mu (H_d^m)}
-\fr{b_a^{H_{+}}}{2\pi }\ln \fr{M_I^2}{\mu (H_{+}^1)\mu (H_{+}^2)}\non \\ 
&-\fr{b_a^D}{2\pi }\ln \fr{M_I^3}{M_{D_1}M_{D_2}M_{D_3}}
-\fr{b_a^L}{2\pi }\ln \fr{M_I^3}{M_{L_1}M_{L_2}M_{L_3}} ~.  
\la{RGEg}
\eeqn 
Here $b_a^D$, $M_D$ and $b_a^L$, $M_L$
denote masses and $b$-factors of $D$ and $L$
states respectively. The matching of $g_1$ and $g_2$
at $M_I$ gives
\beqn
& \ln \fr{M_I}{M_Z}=\fr{5\pi }{18}\l \al_1^{-1}(M_Z)-\al_2^{-1}(M_Z)\r 
+\fr{1}{108}\sum_m\ln \fr{M_I}{\mu (H_d^m)}\non \\
& -\fr{1}{36}\ln \fr{M_I^2}{\mu (H_{+}^1)\mu (H_{+}^2)}
-\fr{1}{27}\ln \fr{M_{L_1}M_{L_2}M_{L_3}}{M_{D_1}M_{D_2}M_{D_3}}~.
\la{MIsc}
\eeqn
At the GUT scale we should have $M_{L_i}=M_{D_i}$.
According to sections \ref{sec:quark} to \ref{sec:nu},  $M_{L_i}\simeq 
M_{D_i}$ should hold
approximately also at lower scales, since they are
determined by $\lan H_3^3\ran $. 
Thus, for the determination of $M_I$ we can safely neglect the last term
in (\ref{MIsc}). Taking the masses
$\mu (H_d^m)\simeq \mu (H_{+}^n)\simeq M_I$, also the second term can be
neglected. 
With
$\al_1^{-1}(M_Z)=59$ and $\al_2^{-1}(M_Z)=29.6$ we then obtain for $M_I$,
the breaking point of the intermediate symmetry,
$M_I\simeq 1.3\cdot 10^{13}$~GeV.
According to our model, however, one extra Higgs $SU(2)_L$ doublet, namely
$(H_A)^{1,2}_2$, should have a mass much below $M_I$, as was discusses in
section \ref{sec:model}. The small VEV found for it, in section 
\ref{sec:quark}, supported this
view. Let us thus take its mass 
$\nu (H_{A2}^{1,2})=M_A=M_{D_3}\approx 4\cdot 10^4$~TeV,
which is far above  the
lowest allowed value ($\sim 500$~TeV) and does not lead to
flavor changing neutral currents. With this value, the second
term in (\ref{MIsc}) leads only to a slight increase of $M_I$:
$M_I\simeq 1.5\cdot 10^{13}$~GeV.
In general, the value of $M_I$ is rather stable with respect to
modifications of our model concerning the Higgs sectors 
$H(\bar 3, 3, 1)$ and $H_A(\bar 3, 3, 1)$.
It is highly interesting that the value obtained for $M_I$ 
is close (see \cite{ft2}) to the 
phenomenologically obtained mass scale 
$(\lam_SF^{\{2,2\}})$ necessary to describe the mass squared difference
observed
in atmospheric neutrino oscillations. Morever, the same scale also
describes the breaking point of the left-right symmetry.

For the precise calculation of $\al_L(M_I)=\al_R(M_I)=\al_{1, 2}(M_I)$
from (\ref{RGEg}), we need input masses for the $D$ quarks and
the leptons $L$. 
The mass of the $3^{\rm rd}$ generation $D$ quark we take is based on the
discussion about an approximate left-right symmetry in the $H$ and $H_A$
sectors. We use
\beq 
M_{D_3}\simeq e_3^3(G^{D\hat{D}})_{33}\simeq 4\cdot 10^4~{\rm TeV}~.
\la{alterII}
\end{equation}
Before renormalization, the lepton $L_3$ has the same mass. The ratios for
the generation splitting of these quarks and leptons are
$\si^4 :\si^2 :1$. The corresponding input in
eq. (\ref{RGEg}) allows now to calculate the values
$\al_3(M_I)$ and $\al_1(M_I)=\al_2(M_I)$, which can then be used as initial
conditions to go up to $M_{\rm GUT}$. After the study of the Yukawa
coupling unification at $M_{\rm GUT}$, one can go back to the scales of
the $D$ and $L$ states to find renormalized values for their masses (see
the next section and the appendix). The corresponding change of
eq. (\ref{RGEg}) will little affect the values of $\al_3$ and
$\al_1=\al_2$ at $M_I$, from which one can start again. The result is
\beqn
&\al_3^{-1}(M_I)=\al_C^{-1}(M_I)\simeq 31.43~,\non \\
&\al_1^{-1}(M_I)=\al_2^{-1}(M_I)=\al_{L,R}^{-1}(M_I)=35.63~.
\la{IalpasMI}
\eeqn
The $D$ and $L$ masses, found this way, are quoted in section 
\ref{sec:desert} and have
already been used in form of the mass matrices 
$M_D=e_3^3G^{D\hat{D}}$
and $M_E=e_3^3G^{L\bar L}$ in sections \ref{sec:quark} and 
\ref{sec:chl}.

Above the scale of $M_I$ $G_{333}$ is unbroken and the quark-lepton
states are unified together with the $D$ states in
$Q_L(3, 1, \bar 3)$, $Q_R(1, \bar 3, 3)$ and the leptons $L$ in $L(\bar 3,
3, 1)$
multiplets.
For the fermion masses we needed besides the VEVs from $H(\bar 3, 3, 1)$
also those from $H_A(\bar 3, 3, 1)$. We take the masses of these Higgses 
to be negligeable for scales above $M_I$ [similar to the mass of 
$H(\bar 3, 3, 1)$]. In fact, we have to do that
because some members lie below $M_I$ and the full $(SU(3))^2$ symmetry
must hold above $M_I$. The corresponding
$b$-factors for $\mu \stackrel{>}{_-}M_I$ are therefore
\beq
\l b_{L}, b_{R}, b_C\r^{M_I} =\l -4, -4, -5\r ~.
\la{b333}
\end{equation}
With these values the meeting point $g_L=g_R=g_C$ would be above the 
Planck scale because $b_L=b_R$ is not much different from $b_C$. 
We know, however, from our
treatment of the charged lepton sector, that the vacuum
expectation values of $H_A(6, 3, 1)$ and $H_A(\bar 3, \bar 6, 1)$
play an important role. Since lying above the $M_I$ scale, the masses of
these two Higgses are equal due to the left-right ${\cal D}_{LR}$
symmetry:
$M(6,3,1)=M(\bar 3,\bar 6,1)\equiv M_6$. 
They contribute to the renormalization with the $b$-factors
\beq
\l b_{L}, b_{R}, b_C\r^6 =\l \fr{7}{2}, \fr{7}{2}, 0\r ~.
\la{b6}
\end{equation}
We now have for $\mu \stackrel{>}{_-}M_I$
\beq
\al_C^{-1}(\mu )=\al_{3}^{-1}(M_I)-\fr{b_{C}^{M_I}}{2\pi }\ln
\fr{\mu }{M_I}~,
\la{alMI3}
\end{equation}
and
\beqn
&\al_{L, R}^{-1}(\mu )=\al_{L, R}^{-1}(M_I)-
\fr{b_{L, R}^{M_I}}{2\pi }\ln\fr{\mu }{M_I}\non \\
&-\te (\mu -M_6)\fr{b_{L, R}^6}{2\pi }\ln\fr{\mu }{M_6}~.
\la{alMILR}
\eeqn
The Grand Unification Energy $M_{\rm GUT}$ can now be obtained by 
setting $\mu =M_{\rm GUT}$ and
equating (\ref{alMI3}) and (\ref{alMILR}). $M_{\rm GUT}$ depends on
$M_6$ and increases with increasing $M_6$.
It is interesting, that even for low values of $M_6$ close to $M_I$
we get a
large values for $M_{\rm GUT}$. For instance for $M_6\simeq 3M_I$ we have
$M_{\rm GUT}\simeq 10^{16}$~GeV.
Already for $M_6\simeq 5\cdot 10^{16}$~GeV
we get $M_{\rm GUT}\simeq 3\cdot 10^{18}$~GeV.
Therefore, in our model we have
$M_{\rm GUT}\stackrel{>}{_\sim}10^{16}$~GeV, which thus insures proton stability
compatible with present experimental limits.
But we still have to see which restrictions are forced on us by
top-bottom-tau unification.

\subsection{Top-bottom-tau unification\label{sec:unifY}}

In this section we study the running of the Yukawa couplings and
their unification. We concentrate on the unification of the third
generation
couplings 
$\lam_t$, $\lam_b$, $\lam_{\tau }$ for the top, bottom and tau fermions,
respectively. In the SM, because of the small mixings in the
quark sector, their evolution is little affected by the other
couplings. 
In the considered model, the situation is different. Apart from the 
fermion
couplings to $H(\bar 3,3,1)$ [first coupling in (\ref{Yukawa})] also
couplings with $H_A$ are important. 
In particular, the Higgses
$H_A^6(6, 3, 1)$, $H_A^{\bar 6}(\bar 3, \bar 6, 1)$ with common mass
$M_6<M_{\rm GUT}$ are important for gauge coupling unification.
Therefore, above the scale $M_I$, the following
Yukawa couplings are relevant for renormalization:
\beqn
&Q_LG_QQ_R~H+\fr{1}{2}LG_LL~H+Q_LA^QQ_R~H_A \non \\
&+\fr{1}{2}LA^LL~H_A^6+\fr{1}{2}L\bar A^LL~H_A^{\bar 6}~.
\la{yuk333}
\eeqn
We have to distinguish the coupling matrices
$G_Q$, $G_L$, $A^Q$, $A^L$, but have $A^L=\bar A^L$
due to the left-right ${\cal D}_{LR}$ symmetry which holds above
$\mu =M_I$. The elements of the diagonal matrices $G_{Q}$, $G_L$ determine
the masses $M_{D_i}$, $M_{L_i}$ respectively.

As a consequence of the first term in (\ref{yuk333})  one has
already at the $G_{333}$ level top-bottom unification: $\lam_t(\mu )$ and
$\lam_b(\mu )$ must
unify at $M_I$ and evolve then further as a single coupling 
$\lam_{Q_3}(\mu )$. This coupling should then unify with 
$\lam_{\tau }(\mu )=\lam_{L_3}(\mu )$ at $\mu =M_{\rm GUT}$. 

Below the scale $M_I$ the coupling matrices $G_Q$, $G_L$, $A^Q$ and
$A^L=\bar A^L$ split into more matrices depending on the Higgs field
components they are attached to. In an obvious notation we have
\beqn
&G_Q\to \l G^{u\hat{u}}~,~~G^{d\hat{d}}~,~~G^{D\hat{D}}\r ~,\non \\
&G_L\to \l G^{e^{-}e^{+}}~,~~G^{L\bar L}~,~~G^{\nu \hat{\nu }}\r ~,\non \\
&A^Q\to \l A^{d\hat{d}},~~A^{d\hat{D}}~,~~A^{D\hat{d}}\r ~,\non \\
&A^L\to \l A^{e^{-}e^{+}}~,~~A^{E^{-}e^{+}}~,~~A^{e^{-}E^{+}}\r ~.
\la{splitGA}
\eeqn 
We left out the matrices $A^{D\hat{D}}$, $A^{E^{-}E^{+}}$ and additional
matrices from the neutral lepton sector. They are multiplied with VEVs
which are -in our model- small compared to competing terms in the same
channel. In the approximations we use for the renormalization
the $G$ matrices remain diagonal and the diagonal elements of the 
matrices $A$ remain zero. Furthermore, the matrices connected to 
$\bar A^L$ are the same as the ones from $A^L$. But the matrices
derived from 
$A^L=\bar A^L$ are no more strictly antisymmetric.

The most important elements of the matrices (\ref{splitGA}) are the
$(3,3)$ elements of the $G$'s and the $(2,3)$ and $(3,2)$ elements of the
$A$'s: $(G^{u\hat{u}})_{33}=\lam_t(\mu )$, 
$(G^{e^{-}e^{+}})_{33}=\lam_{\tau }(\mu )$ etc.
For the matrix elements of $A^{d\hat{d}}$ we define
\beq
(A^{d\hat{d}})_{23}={\rm i}\ov{\lam }_A(\mu )~,~~~~
(A^{d\hat{d}})_{32}=-{\rm i}\hat{\lam }_A(\mu )~.
\la{deflamA}
\end{equation}
Clearly, we have $\ov{\lam }_A(\mu )=\hat{\lam }_A(\mu )$ for 
$\mu \stackrel{>}{_-}M_I$.

There is a restriction from the mass of the vector boson $W$ for a
combination of the VEVs multiplying the coupling matrices. With the
notation
\beq
e^1_1=v_0\sin \bt ~,~~~~~
e^2_2=v_0\cos \bt ~,
\la{notV0}
\end{equation}
the condition is
\beqn
&v_0^2+\l f^2_2\r^2 
+\l f^2_3\r^2+\l f_{\{1,3\}3}\r^2 \non \\
&+\l f^{2\{13\}}-f_{\{13\}2}\r^2=
(174~{\rm GeV})^2~.
\la{VEVcond}
\eeqn
Since $e_1^1$ and $e_2^2$ contribute to the masses of the third
generation, the $v_0^2$ term should be the dominant one.
At the scale $\mu =M_Z$ one has
\beqn
&\lam_t(M_Z)=\fr{m_t}{v_0\sin \bt }~,~~~~
\lam_b(M_Z)=\fr{m_b^0}{v_0\cos \bt }~,\non \\
&\lam_{\tau }(M_Z)=\fr{m_{\tau }^0}{v_0\cos \bt }~.
\la{tbtau}
\eeqn
%
Here $m_b^0$ and $m_{\tau }^0$ are a little smaller than $m_b$ and
$m_{\tau }$ respectively, since they refer to the diagonal parts of the
down quark and
charged lepton mass matrices. In sections \ref{sec:quark} and 
\ref{sec:chl}, we found 
$m_b^0/m_b\simeq 0.989$, $m_{\tau }^0/m_{\tau }\simeq 0.966$.

We can now set up the renormalization group equations for
$\lam_t$, $\lam_b$, $\lam_{\tau }$, $\ov{\lam }_A$ and
$\hat{\lam }_A$. They are connected with each other and - due to the 
$SU(3)_L\tm SU(3)_R$ symmetry at $\mu \stackrel{>}{_-}M_I$ - no other
coupling intervenes.
Below $M_I$ we have for 
$\eta_t=\fr{\lam_t^2}{4\pi }$, $\eta_b=\fr{\lam_b^2}{4\pi }$, 
$\eta_{\tau }=\fr{\lam_{\tau }^2}{4\pi }$,  
$\hat{\eta }_A=\fr{(\hat{\lam }_A)^2}{4\pi }$ and 
$\ov{\eta }_A=\fr{(\ov{\lam }_A)^2}{4\pi }$
\bwi
\beq
2\pi
\eta_t'=\fr{9}{2}\eta_t^2+\fr{1}{2}\eta_t\eta_b-
\eta_t\l \fr{17}{20}\al_1+\fr{9}{4}\al_2+8\al_3\r
+\te (\mu -M_A) \fr{1}{2}\eta_t\hat{\eta }_A ~,
\la{RGtop}
\end{equation}
\beq
2\pi
\eta_b'=\fr{9}{2}\eta_b^2+\fr{1}{2}\eta_b\eta_t+\eta_b\eta_{\tau }-
\eta_b\l \fr{1}{4}\al_1+\fr{9}{4}\al_2+8\al_3\r
+\te (\mu -M_A)\eta_b \l \fr{1}{2}\hat{\eta }_A+\ov{\eta }_A\r ~~,
\la{RGbot}
\end{equation}
\beq
2\pi
\eta_{\tau }'=\fr{5}{2}\eta_{\tau }^2+3\eta_{\tau }\eta_b
-\eta_{\tau }\l \fr{9}{4}\al_1+\fr{9}{4}\al_2\r ~,
\la{RGtau}
\end{equation}
\beq
2\pi
\hat{\eta }_A'=\fr{1}{2}\hat{\eta }_A\l \eta_b
+\eta_t -\fr{1}{5}\al_1-3\al_2-16\al_3\r 
+\te (\mu -M_A) \hat{\eta }_A\l \fr{9}{2}\hat{\eta }_A
+3\ov{\eta }_A-\fr{3}{20}\al_1 -\fr{3}{4}\al_2\r~,
\la{RGhatA}
\end{equation}
\beq
2\pi \hs{0.3mm}
{\ov{\eta }_A}'=\ov{\eta }_A\l \eta_b
-\fr{1}{10}\al_1-\fr{3}{2}\al_2-8\al_3\r 
+\te (\mu -M_A) \ov{\eta }_A\l \fr{9}{2}\ov{\eta }_A
+3\hat{\eta }_A-\fr{3}{20}\al_1 -\fr{3}{4}\al_2\r~.
\la{RGovA}
\end{equation} 
\ewi 
At $\mu =M_I$ the matching
\beq
\eta_t=\eta_b\equiv \eta_{Q_3}~,~~~
\eta_{\tau }\equiv \eta_{L_3}~,~~~
\ov{\eta }_A=\hat{\eta }_A \equiv \eta_A~,
\la{matchMI}
\end{equation}
is required.

Above $M_I$ we have for $\eta_{Q_3}$, $\eta_{L_3}$, $\eta_A$ and
$\eta_{AL}=\fr{(\lam^L_A)^2}{4\pi }$ the equations
\beq
2\pi
\eta_{Q_3}'=6\eta_{Q_3}^2+\eta_{Q_3}\eta_{L_3}+
3\eta_{Q_3}\eta_A
-\eta_{Q_3}8\l \al_{L, R}+\al_C\r~,
\la{RGbotMI}
\end{equation}
\beqn
&2\pi \eta_{L_3}'=2\eta_{L_3}^2+3\eta_{L_3}\eta_{Q_3}
-\eta_{L_3}\fr{56}{3}\al_{L, R}\non \\
&+\te (\mu -M_6)3\eta_{L_3}\eta_{AL}~,
\la{RGtauMI}
\eeqn
\beq
2\pi \eta_A'=9\eta_A^2+\fr{3}{2}\eta_A\eta_{Q_3}-
\eta_A 8\l \al_{L, R}+\al_C\r ~,
\la{RGAMI}
\end{equation}
\beqn
&2\pi \eta_{AL}'=\eta_{AL}\l \fr{1}{2}\eta_{L_3}-16\al_{L, R}\r +\non \\
&\te (\mu -M_6) \eta_{AL}\l 4\eta_{AL}-\fr{14}{3}\al_{L, R}\r ~.
\la{RGSMI}
\eeqn
${\rm i}\lam_A^L$ is the $(2,3)$ element of $A^L=\bar A^L$ and is only 
needed above $M_I$.
The matching condition at $M_{\rm GUT}$ for the final unification of the
couplings reads
\beq
\eta_{Q_3}(M_{\rm GUT})=\eta_{L_3}(M_{\rm GUT})~,~~
\eta_A(M_{\rm GUT})=\eta_{AL}(M_{\rm GUT})~.
\la{matchMG}
\end{equation}

The procedure of finding a solution with gauge and top-bottom-tau
unification is the following: A given value of 
$M_{\rm GUT}\stackrel{>}{_\sim }10^{16}$~GeV (otherwise no solution is
possible) fixes $M_6$. Taking then trial values for
$\eta_{Q_3}(M_{\rm GUT})$ and $\eta_A(M_{\rm GUT})$ and solving eqs.
(\ref{RGbotMI})-(\ref{RGSMI}) gives their values at $M_I$. These values
determine $\eta_t(M_I)=\eta_b(M_I)$, $\eta_{\tau }(M_I)$ and 
$\ov{\eta }_A(M_I)=\hat{\eta }_A(M_I)$. The renormalization
group equations (\ref{RGtop})-(\ref{RGovA}) allow then to calculate
$\lam_t(M_Z)$, $\lam_b(M_Z)$, $\lam_{\tau }(M_Z)$. Clearly, the input
values $\eta_{Q_3}(M_{\rm GUT})$ and $\eta_A(M_{\rm GUT})$ have now to be
changed such that $\lam_{\tau }/\lam_b$ becomes equal to 
$m_{\tau }^0/m_b^0$ and $\lam_t$, $\lam_b$ are in the perturbative
region i.e. $\stackrel{<}{_\sim }3$. If this can be achieved, one can
calculate from (\ref{tbtau}) $v_0^2$ and $\tan \bt $
\beq
v_0^2=\fr{m_t^2}{\lam_t^2}+\fr{(m_b^0)^2}{\lam_b^2}~,~~~
\tan \bt =\fr{m_t}{m_b^0}\fr{\lam_b}{\lam_t}~.
\la{v0tg}
\end{equation}
Of course, only solutions with $v_0< v=174$~GeV are acceptable.

%
%
\begin{figure}
\rput(0.3,-0.5){\large 'Concorde'}
\begin{center}
\leavevmode
\leavevmode
\vspace{2.5cm}
\includegraphics{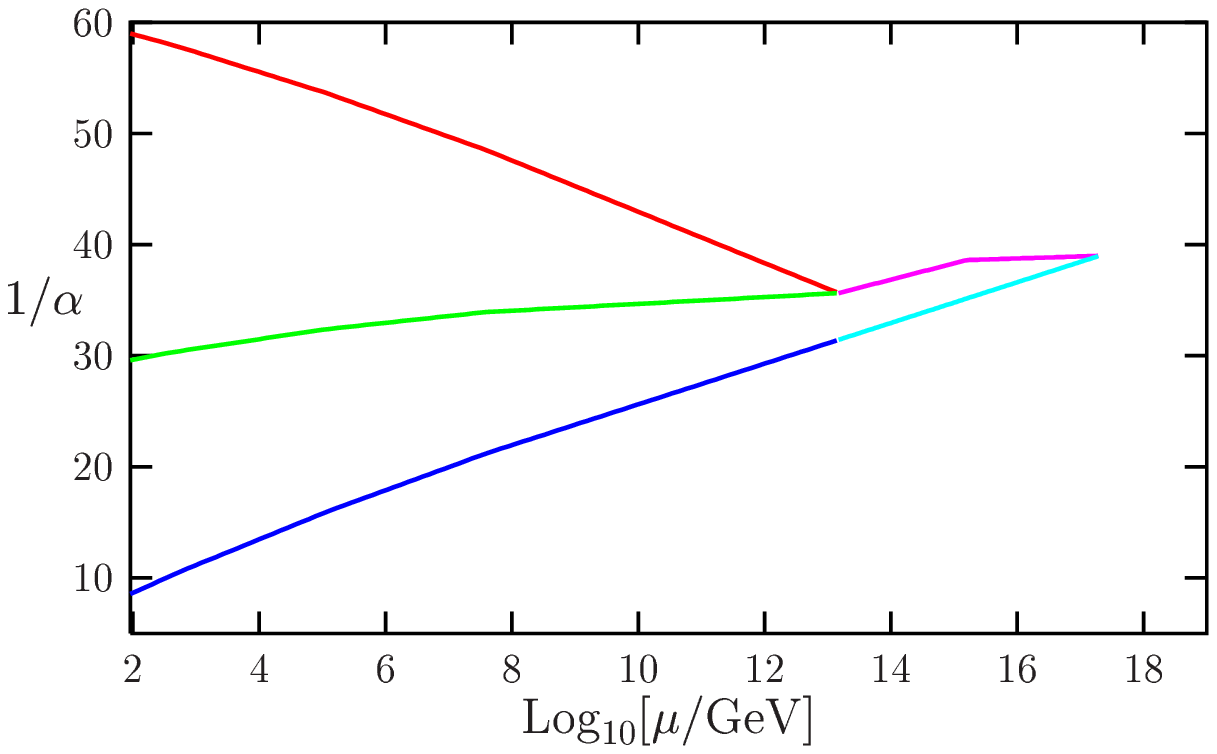}
\end{center}
\vs{3cm}
\caption{
Unification of gauge couplings.
$M_I\simeq 1.5\cdot 10^{13}$~GeV, $M_6\simeq 1.6\cdot 10^{16}$~GeV,
$M_{\rm GUT}\simeq 2\cdot 10^{17}$~GeV and $\al_G^{-1}\simeq 39$.
}
\end{figure}
%
\subsection{Numerical solution for \\
$M_{\rm GUT}=2\cdot 10^{17}~{\rm GeV}$,
$M_{D_3}=M_A=4\cdot 10^4~{\rm TeV}$\label{sec:unifNum}}

Here we present a numerical solution of the problem of gauge and Yukawa
coupling unification in $E_6$, which satisfies all above mentioned
requirements. We choose the unification scale to be 
$2\cdot 10^{17}$~GeV, the
masses of the heaviest $D$ state and 
the Higgs field $(H_A)_2^{1,2}$ both equal to
$3.98\cdot 10^7$~GeV.

Further imput values are the third generation masses
\beqn
&m_t(M_Z)=173~{\rm GeV}~,~~~~
m_b^0(M_Z)=2.859~{\rm GeV}~,\non \\
&m_{\tau }^0(M_Z)=1.689~{\rm GeV}~,
\la{inputFer}
\eeqn
the three gauge coupling constants at $\mu =M_Z$ and a suitable value for
$\eta_{t, b, \tau }$ at $M_{\rm GUT}$
\beq
\eta_{t, b, \tau }(M_{\rm GUT})\simeq 0.0381~.
\la{suitbtau}
\end{equation}
{}For this latter value all couplings remain in the perturbative region
and $v_0<v=174$~GeV.
As a result we find the solution
\beqn
&M_6\simeq 1.6\cdot 10^{16}~{\rm GeV}~,~~~~~
M_I\simeq 1.5\cdot 10^{13}~{\rm GeV}~,\non \\
&\eta_A(M_{\rm GUT})=\eta_{AL}(M_{\rm GUT})\simeq 0.0336~,
\la{MAMIYGUTmod}
\eeqn
with the following consequences:
\beqn
&\al_G^{-1}(M_{\rm GUT})\simeq 38.99~,\non \\
&\eta_{t, b}(M_I)=\eta_{Q_3}(M_I)\simeq 0.0412~,\non \\
&\eta_{\tau }(M_I)=\eta_{L_3}(M_I)\simeq 0.0536~,\non \\
&\hat{\eta }_A(M_I)=\ov{\eta }_A(M_I)=\eta_A(M_I)\simeq 0.0368~,\non \\
&\eta_{AL}(M_I)\simeq 0.0598~,
\la{alGYMImod}
\eeqn
\beqn  
&\lam_{\tau }(M_Z)=\lam_b(M_Z)\fr{m_{\tau }^0}{m_b^0}\simeq 0.612~,~~~~~~
\lam_t(M_Z)\simeq 1.127~,\non \\
&v_0\simeq 153.48~{\rm GeV}~,~~~~~~
\tan \bt \simeq 55.59~.
\la{v0tgmod}
\eeqn
{}From the value found for $v_0$, we can now determine
$(f^2_3)^2+(f_{\{13\}3})^2$ from (\ref{VEVcond}).
Using then (\ref{fixf2f3e33}), (\ref{inpLep}) together with 
$e_3^3(G^{D\hat{D}})_{33}=3.98\cdot 10^7$~GeV and 
$|f^2_3|=|f^3_2|$, $|f_{\{13\}3}|=|f^{3\{13\}}|$, 
we finally get
\beqn
&f_3^2=\pm 43.504~{\rm GeV}~,~~~f_2^3=\mp 43.504~{\rm GeV}~,\non \\
&f_{\{13\}3}=f^{3\{13\}}=69.484~{\rm GeV}~.
\la{VEVsmod}
\eeqn 
%
%
\begin{figure}
\rput(0.4,-0.5){\large 'Bermuda Triangle'}
\rput(-2.15,-2.6){\large $\eta_{t}$}
\rput(-2.6,-3.6){\large $\eta_{b}$}
\rput(-2,-4.7){\large $\eta_{\tau }$}
\begin{center}
\leavevmode
\leavevmode
\vspace{2.5cm}
\includegraphics{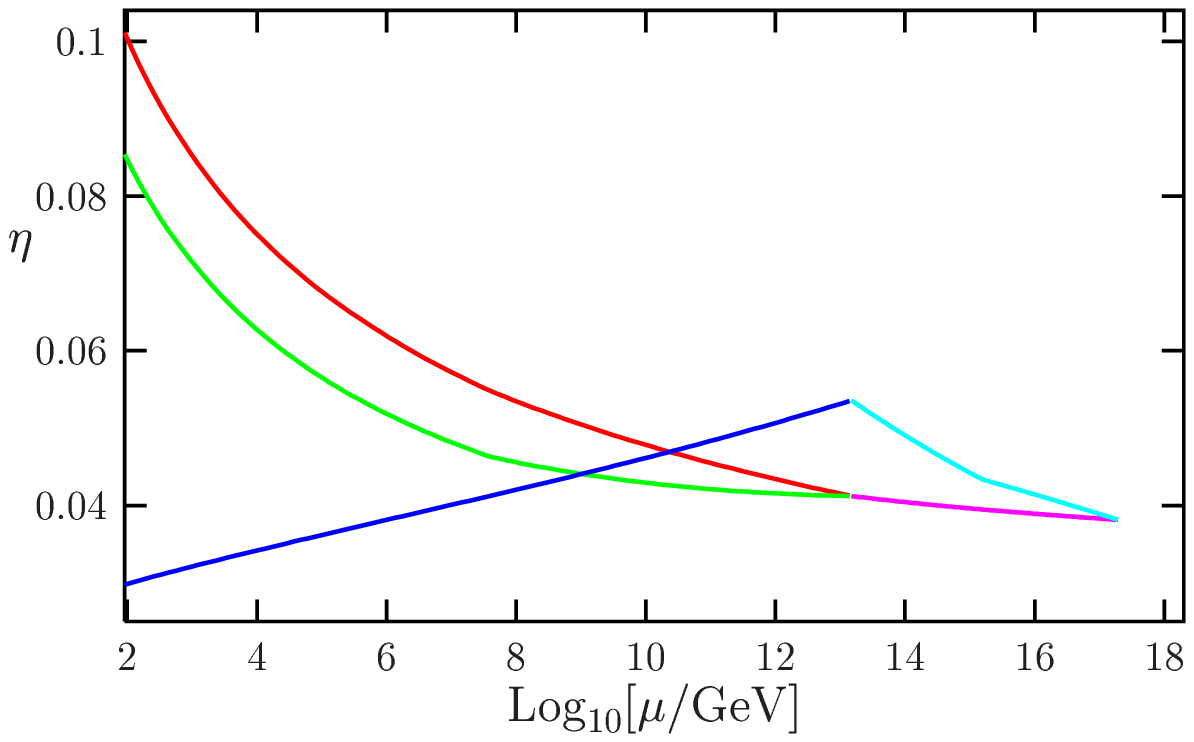}
\end{center}
\vs{3cm}
\caption{
$t-b-\tau $ unification:
$\lam_t(M_{\rm GUT})=\lam_b(M_{\rm GUT})=\lam_{\tau }(M_{\rm GUT})
\simeq 0.692$.
}
\end{figure}
%
%
%
%
\begin{figure}
\rput(0.4,-0.5){\large 'Desert Spider'}
\rput(2.9,-4.5){\large $\eta_{AL}$}
\rput(2,-4.9){\large $\eta_{A}$}
\begin{center}
\leavevmode
\leavevmode
\vspace{2.5cm}
\includegraphics{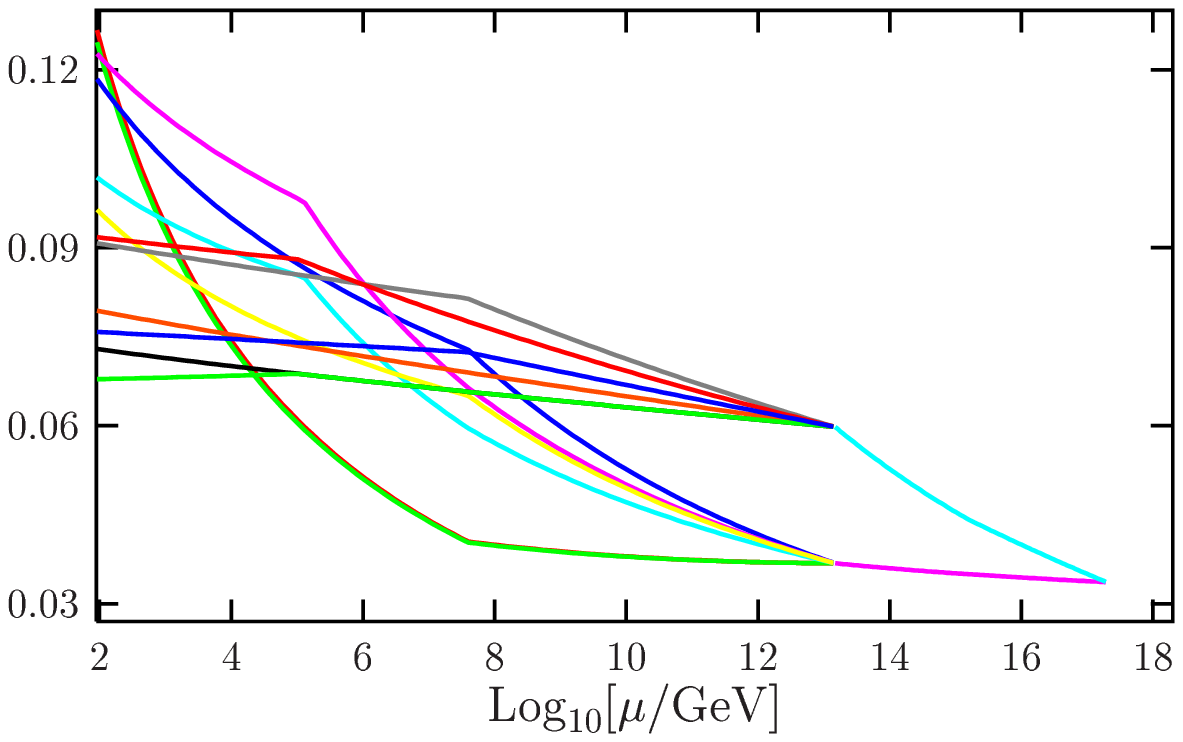}
\end{center}
\vs{3cm}
\caption{ 
Unification of (2,3), (3,2) elements of $A^Q$ and $A^L$ matrices.
$A^Q_{23}(M_{\rm GUT})=A^L_{23}(M_{\rm GUT})\simeq 0.65\hs{0.5mm}{\rm i}$.
}
\end{figure}
The solution for the gauge coupling and Yukawa coupling unification given
here has been applied in the previous sections, in particular, for the
evaluation of the renormalization parameters for all the different mass
matrices.

In figure 1 -'Concorde'- we show the evolution of the gauge couplings and
their
unification.
Figure 2  - 'Bermuda triangle'- exhibits the running of the
Yukawa couplings $\eta_t$, $\eta_b$, $\eta_{\tau }$ and their unification.
In figure 3 - 'desert spider' -  the running of the (2,3) and
(3,2) elements
of the $A$-matrices and their unification is presented. 
In these evaluations the splittings
between the masses $M_{D_i}$ and $M_{L_i}$ (which we discuss in an
appendix)
have been taken into account.

\section{Conclusions\label{sec:conc}}

The $E_6$ model presented has many attractive features. Only few input
data are sufficient to obtain a realistic picture of the fermion
masses and their mixings. The presence of new heavy fermions in the
'desert' plays an important role even for the mass matrices of the SM
particles. All generation mixings and ${\cal CP}$ violations arise from a 
single antisymmetric matrix $A$, which mixes the light fermions but also 
the light with the
heavy fermions. The latter effect also contributes in an important
way to the eigenvalues of the quark and lepton mass matrices. 
For instance, the main part of the $\mu $ meson mass and of the strange 
quark mass is generated by virtual transitions to heavy fermions.
As a side remark we note, that the antisymmetric generation
mixing matrix found here could lead to significant effects in rare weak 
decay processes with fixed phases of the new contributions. 
The matrix $A$, in combination
with $G$, is also responsible for the bilarge mixing of the light
neutrinos and their oscillation pattern. In the limit of no
renormalization effects, the neutrino mixing is bimaximal.

Those heavy new particles, which form  $10$-plets with respect to
$SO(10)$, have a hierarchical spectrum similar to the spectrum of the up
quarks. The lightest ones are expected to lie in the low TeV region.

The group $E_6$ provides new insights about the unification of the three
gauge couplings and about the unification of the Yukawa couplings of top,
bottom and tau. 
The intermediate symmetry $SU(3)_L\tm SU(3)_R\tm SU(3)_C$
with a discrete left-right symmetry plays a decisive
role. The breaking point of this intermediate symmetry is fixed by the
known gauge couplings $g_1$ and $g_2$. 
Simultaneously it determines the mass scales for the light and heavy
neutrinos.
We achieved a solution of the gauge
and Yukawa coupling unification with strongly constraint parameters. It
describes the evolution and the final convergence of many coupling
matrices which differ significantly at low energies. The solution allows
to calculate quite a number of properties such as transition matrices from
heavy to light fermions, Majorana phases and the double $\bt $-decay
matrix element.
Due to the high unification scale 
($> 10^{16}$~GeV), the model adequately 
suppresses dimension six operators which induce nucleon decays. The proton
lifetime is above the presently accessible range.

The presented $E_6$ model can
be supersymmetrized without changing the construction of the Yukawa
sector. A
supersymmetric version would, however, affect the coupling unification
picture given here.

\begin{acknowledgments}
We thank Qaisar Shafi for interesting discussions and Mathias
Jamin for providing us with the newest data on the quark masses.
\end{acknowledgments}

\section*{APPENDIX: Renormalization Analysis}
\setcounter{equation}{0}
\renewcommand{\theequation}{A.\arabic{equation}}   

\subsection*{A1.~The $D$ and $L$ masses}

In our model the $D$ and $L$ states play a crucial role for the
light families. 
Also, the mass splitting between these states must be taken into
account
in the study of the gauge coupling unification.
So let's start with the renormalization of their masses. 
The Yukawa interactions occuring in
(\ref{yuk333}) obey the boundary
condition $G=G_Q=G_L$ at $\mu =M_{\rm GUT}$ due to $E_6$ symmetry.
We write
\beq
G_Q={\rm Diag}\l \lam_{Q_1}\si^4~,~-\lam_{Q_2}\si^2~,~\lam_{Q_3}\r~,
\la{GQ}
\end{equation}
and
\beq
G_L={\rm Diag}\l \lam_{L_1}\si^4~,~-\lam_{L_2}\si^2~,~\lam_{L_3}\r~.
\la{GL}
\end{equation}
{}For a given value for $\lam_Q=\lam_L$ at $M_{\rm GUT}$ the
values for $\lam_{Q_3}$, $\lam_{L_3}$ at $\mu =M_I$ can be
obtained using eqs. (\ref{RGbotMI})-(\ref{RGSMI}).
%
%
The gauge interactions contribute to the running of $\lam_{Q_{1(2)}}$ and
$\lam_{L_{1(2)}}$ in a similar way as for 
$\lam_{Q_3}$ and $\lam_{L_3}$. However,
also the Yukawa interactions are to be taken into account. 
One has
\beqn
&\left. \fr{\lam_{Q_1}}{\lam_{Q_3}}\right |_{M_I}=
\left. \fr{\lam_{Q_1}}{\lam_{Q_3}}\right |_{M_{\rm 
GUT}}P_{Q_3}^3P_A^3~,\non \\
&\left. \fr{\lam_{Q_2}}{\lam_{Q_3}}\right |_{M_I}=
\left. \fr{\lam_{Q_2}}{\lam_{Q_3}}\right |_{M_{\rm GUT}}P_{Q_3}^3~,
\la{Q12Q3MI}
\eeqn
\beqn
&\left. \fr{\lam_{L_1}}{\lam_{L_3}}\right |_{M_I}=
\left. \fr{\lam_{L_1}}{\lam_{L_3}}\right |_{M_{\rm 
GUT}}P_{L_3}P_{AL}^3~,\non \\
&\left. \fr{\lam_{L_2}}{\lam_{L_3}}\right |_{M_I}=
\left. \fr{\lam_{L_2}}{\lam_{L_3}}\right |_{M_{\rm GUT}}P_{L_3}~,
\la{L12L3MI}
\eeqn 
where the $P$-factors are defined as follows
\beqn
&P_{Q_3(L_3, A)}=\exp \lq \fr{1}{4\pi }\int_{M_I}^{M_{\rm GUT}}
\hs{-0.2cm}\eta_{Q_3(L_3, A)}(\mu')d\ln \mu' \rq ~,\non \\
&P_{AL}=\exp \lq \fr{1}{4\pi }\int_{M_6}^{M_{\rm GUT}}  
\hs{-0.2cm}\eta_{AL}(\mu')d\ln \mu' \rq ~.
\la{Pfactors}
\eeqn
All numerical analysis are carried out  according to the
solution presented in section \ref{sec:unifNum}.
The $P$-factors have the values
\beqn 
&P_{Q_3}\simeq 1.030~,~~~P_{L_3}\simeq 1.034~,~\non \\
&P_A\simeq 1.027~,~~~~P_{AL}\simeq 1.015~.
\la{numPfact}
\eeqn
At the scale $M_I$ the matrices $G_Q$ and $G_L$
are given by
\beq
G_Q(M_I)=
{\rm Diag}\l \ka_1^Q\si^4,~-\ka_2^Q\si^2,~1\r \lam_{Q_3}(M_I)~,
\la{YQMI}
\end{equation}
\beq
G_L(M_I)=
{\rm Diag}\l \ka_1^L\si^4,~-\ka_2^L\si^2,~1\r \lam_{L_3}(M_I)~,
\la{YLMI}
\end{equation}
where
\beqn
&\ka_1^Q=\fr{1}{\ka_u}P_{Q_3}^3P_A^3~,~~~
\ka_2^Q=\fr{1}{\ka_c}P_{Q_3}^3~,\non \\
&\ka_1^L=\fr{1}{\ka_u}P_{L_3}P_{AL}^3~,~~~
\ka_2^L=\fr{1}{\ka_c}P_{L_3}~.
\la{kapQL}
\eeqn
The coefficients $\ka_u$ and $\ka_c$ are the factors of $\si^4$
and $\si^2$ occuring in $G$ at the GUT scale [see (\ref{GatMG})]
as obtained from $m_U$ at $\mu =M_Z$ [eq. (\ref{upqMz})].
(\ref{YQMI}) and (\ref{YLMI}) allow to determine at $M_I$ the ratios 
$M_{D_i}/M_{D_3}$ and $M_{L_i}/M_{L_3}$ since $M_D\sim G_Q$, $M_L\sim
G_L$. 
%
Below $M_I$ the corresponding ratios run due to
gauge interactions. We have
\beqn
&\left. \fr{\lam_{D_1}}{\lam_{D_3}}\right |_{M_{D_1}}=
\left. \fr{\lam_{D_1}}{\lam_{D_3}}\right |_{M_I}\non \\
&\tm \rho_{\al_1}^{2/5}(M_{D_1})\rho_{\al_3}^8(M_{D_1})
\rho_{\al_1}^{-2/5}(M_{D_3})\rho_{\al_3}^{-8}(M_{D_3})~,\non \\
&\left. \fr{\lam_{D_2}}{\lam_{D_3}}\right |_{M_{D_2}}=
\left. \fr{\lam_{D_2}}{\lam_{D_3}}\right |_{M_I}\non \\
&\tm \rho_{\al_1}^{2/5}(M_{D_2})\rho_{\al_3}^8(M_{D_2})
\rho_{\al_1}^{-2/5}(M_{D_3})\rho_{\al_3}^{-8}(M_{D_3})~.
\la{lamDiD3}
\eeqn
The scaling factors 
$\rho_i$ are defined in (\ref{intrho}) in a similar way as the
$P$-factors. In our model we find
\beqn 
&\rho_{\al_1}(M_{D_2})\simeq 1.034~,~~
\rho_{\al_3}(M_{D_2})\simeq 1.066~,\non \\
&\rho_{\al_1}(M_{D_3})\simeq 1.025~,~~
\rho_{\al_3}(M_{D_3})\simeq 1.040~.
\la{ro13MD23}
\eeqn 
The ratios (\ref{lamDiD3})
do not run below the scales $\mu =M_{D_1}$ and $M_{D_2}$ 
respectively.
Combining (\ref{YQMI}) and (\ref{lamDiD3}),  we get
for the mass ratios of the $D$-states
\beq
M_{D\hat{D}}=e_3^3G^{D\hat{D}}={\rm Diag}\l \ka^{D\hat{D}}_1\si^4,~
-\ka^{D\hat{D}}_2\si^2,~1\r M_{D_3} ~,
\la{MDDspace}
\end{equation}
\beq
\ka^{D\hat{D}}_1\simeq 1.237~,~~~~~~~~
\ka^{D\hat{D}}_2\simeq 0.965~.
\la{kapaDD}
\end{equation}

{}For the mass ratios of heavy quarks to heavy leptons we have at 
$\mu =M_I$ 
\beqn
&\left. \fr{\lam_{Q_1}}{\lam_{L_1}}\right |_{M_I}=
\left. \fr{\lam_{Q_3}}{\lam_{L_3}}\right |_{M_I}
P_{Q_3}^3P_A^3P_{L_3}^{-1}P_{AL}^{-3}~,\non \\
&\left. \fr{\lam_{Q_2}}{\lam_{L_2}}\right |_{M_I}=
\left. \fr{\lam_{Q_3}}{\lam_{L_3}}\right |_{M_I}
P_{Q_3}^3P_{L_3}^{-1}~.
\la{Q1L1Q2L2}
\eeqn 
Using (\ref{alGYMImod}) and (\ref{Pfactors}),  (\ref{Q1L1Q2L2})
gives 
\beq
\left. \fr{M_{D_1}}{M_{L_1}}\right |_{M_I}=0.960~,~~~~~~
\left. \fr{M_{D_2}}{M_{L_2}}\right |_{M_I}=0.928~.
\la{ratiosMI}
\end{equation}
These values form the starting point for the further running of
these mass ratios down to their own mass scale due to their gauge
interactions
\beq
\left. \fr{M_{D_i}}{M_{L_i}}\right |_{\mu =M_{D_i} } =     
\left. \fr{M_{D_i}}{M_{L_i}}\right |_{\mu =M_I}
\prod_{a=1}^{3}\l \fr{\al_a(M_I)}{\al_a(M_{D_i})}\r 
^{\fr{\tl{c}_a^D-\tl{c}_a^L}{2\tl{b}_a}}~.
\la{relDL}  
\end{equation}  
$\tl{b}_a$ are the $b$-factors in the interval $M_{D_i}-M_I$ and
\beqn 
&\tl{c}_a^D=\l \fr{2}{5}, 0, 8\r ~,~~~~  
\tl{c}_a^L=\l \fr{9}{10}, \fr{9}{2}, 0\r ~,\non \\
&a=(U(1)_Y, SU(2)_L, SU(3)_C)~.
\la{tlcDL}   
\eeqn
The heavy quark to heavy lepton mass ratios at their mass scales [see
(\ref{alterII}),
(\ref{MDDspace}), (\ref{kapaDD})]
are found to be
\beq 
\fr{M_{D_1}}{M_{L_1}}=1.572 ~,~~
\fr{M_{D_2}}{M_{L_2}}=1.249 ~,~~ 
\fr{M_{D_3}}{M_{L_3}}=1.039 ~. 
\la{RATcases} 
\end{equation}   
These mass splittings do only minimally  affect the gauge coupling
unification. Their effect is a subleading one only.
By combining  (\ref{kapaDD}) and (\ref{RATcases}), one can find
the mass ratios for the $L$-states
\beq
M_{L\bar L}=e_3^3G^{L\bar L}={\rm Diag}\l \ka^{L\bar L}_1\si^4,~
-\ka^{L\bar L}_2\si^2,~1\r M_{L_3} ~,
\la{MLLspace}
\end{equation}
with
\beq
\ka^{L\bar L}_1\simeq 0.817~,~~~~~~~~~
\ka^{L\bar L}_2\simeq 0.802.
\la{kapaLL}
\end{equation}
Recall that in the application of our model we
use $M_{D_3}=3.98\cdot 10^7$~GeV.

\subsection*{A2.~The running of the matrices~ 
$G^{u\hat{u}}$, $G^{d\hat{d}}$~ and~ $G^{e^{-}e^+}$}

Below the scale $M_I$ instead of the matrix $G_Q$ we have two matrices: 
$G^{u\hat{u}}$ and $G^{d\hat{d}}$. 
$G^{u\hat{u}}$ accounts for the up quark Yukawa couplings.
At $\mu =M_Z$ it is  given by (\ref{upq}), (\ref{upqMz}). Its
hierarchical structure is dictated from the values of up-type quark masses
(\ref{expQLmassMZ}). The diagonal matrix
$G^{d\hat{d}}$ describes the coupling of $q\hat{d}$ with $H_2^{1,2}$. 
The matrix $G_L$ is also splitted below $\mu =M_I$, namely to 
$G^{e^{-}e^+}$ and $G^{\nu \hat{\nu }}$, which describe the
couplings $le^{+}H_2^{1,2}$ and
$l\hat{\nu }H_1^{1,2}$, respectively. At the scale $M_I$ 
these couplings satisfy the conditions
\beqn
&G^{u\hat{u}}(M_I)=G^{d\hat{d}}(M_I)=G_Q(M_I)~,\non \\
&G^{e^{-}e^+}(M_I)=G^{\nu \hat{\nu }}(M_I)=G_L(M_I)~.
\la{uncondMI}
\eeqn
Since we have explicit information on $G^{u\hat{u}}$ at $M_Z$, we can 
calculate  $G$ at the GUT
scale and then derive $G^{d\hat{d}}$, $G^{e^{-}e^+}$ at 
$\mu =M_Z$. $G^{\nu \hat{\nu }}$
is connected with the neutrino sector, the renormalization of which 
is studied in appendix A4.

Below $M_I$ the running of the up and charm quark Yukawa couplings
are due to following equations
\beq
4\pi \lam_u'=\lam_u \l 3\eta_t-\fr{17}{20}\al_1-\fr{9}{4}\al_2-8\al_3\r ~,
\la{uqRG}
\end{equation}
\beq
4\pi \lam_c'=\lam_c \l 3\eta_t-\fr{17}{20}\al_1-\fr{9}{4}\al_2-8\al_3
+\te (\mu -M_A)\fr{1}{2}\ov{\eta }_A\r ,
\la{cqRG}
\end{equation}
It is convenient to introduce for each coupling 
$\eta_i$ ($\eta_t, \al_1, \al_2, \cdots $) the quantity
\beq
\rho_i(\mu )=\exp \lq \fr{1}{4\pi }\int_{\mu }^{M_I}
\eta_i(\mu' )d\ln \mu' \rq .
\la{deffactor}
\end{equation}
These scaling factors allow to express the coupling constants at arbitrary
scales and thus are useful for satisfying the matching conditions on
boundaries.
According to (\ref{deffactor}), we also have
\beq
\exp \lq \fr{1}{4\pi }\int_{\tl{\mu_1 }}^{\tl{\mu_2 }}  
\eta_i(\mu' )d\ln \mu' \rq =\rho_i(\tl{\mu_1 })/\rho_i(\tl{\mu_2 })~.
\la{intrho}
\end{equation}  
Taking all this into account, equations
(\ref{RGtop}), (\ref{uqRG}), (\ref{cqRG}) give
\beq
\left. \fr{\lam_u}{\lam_t}\right|_{M_Z}=
\left. \fr{\lam_u}{\lam_t}\right|_{M_I}
\rho_t^{3/2}(M_Z)\rho_b^{1/2}(M_Z)\rho_{\hat{\eta }_A}^{1/2}(M_A)~, 
\la{UratMZ}
\end{equation}
\beq
\left. \fr{\lam_c}{\lam_t}\right|_{M_Z}=
\left. \fr{\lam_c}{\lam_t}\right|_{M_I}
\rho_t^{3/2}(M_Z)\rho_b^{1/2}(M_Z)
\rho_{\hat{\eta }_A}^{1/2}(M_A)\rho_{\ov{\eta }_A}^{-1/2}(M_A). 
\la{CratMZ}
\end{equation}
At the scale $M_I$ the ratios $\lam_{u,c}/\lam_t$ can be
expressed by $\left.\fr{\lam_{Q_{1,2}}}{\lam_{Q_3}}\right|_{M_{\rm GUT}}$
according to (\ref{Q12Q3MI}). Therefore, we have
\beq
\left. \fr{\lam_u}{\lam_t}\right|_{M_Z}=   
\left. \fr{\lam_u}{\lam_t}\right|_{M_{\rm GUT}}\kappa_u~,~~~~
\left. \fr{\lam_c}{\lam_t}\right|_{M_Z}=   
\left. \fr{\lam_c}{\lam_t}\right|_{M_{\rm GUT}}\kappa_c~,  
\la{UCratMZMG}
\end{equation}
where
\beqn
&\kappa_u =P_{Q_3}^3P_{A}^3
\rho_t^{3/2}(M_Z)\rho_b^{1/2}(M_Z)\rho_{\hat{\eta }_A}^{1/2}(M_A)~,\non \\
&\kappa_c=\kappa_u P_{A}^{-3}\rho_{\ov{\eta }_A}^{-1/2}(M_A)~. 
\la{kapaUC}
\eeqn 
Numerically we find in our specific model
\beqn
&\rho_t(M_Z)\simeq 1.130~,~~~
\rho_b(M_Z)\simeq 1.112~,\non \\
&\rho_{\hat{\eta }_A}(M_A)\simeq \rho_{\ov{\eta }_A}(M_A)\simeq 1.040~.
\la{rotbA}
\eeqn
Consequently, if $G$ has at $\mu =M_Z$ the form given in
(\ref{upqMz}), we have to take at the GUT scale
\beq
G(M_{\rm GUT})={\rm Diag }\l \fr{1}{\kappa_u}\si^4 ,~
-\fr{1}{\kappa_c}\si^2 ,~ 1\r \cdot \lam_t(M_{\rm GUT})~.
\la{GatMG}
\end{equation}
Our model gives 
\beq
\kappa_u\simeq 1.528~,~~~~~\kappa_c\simeq 1.386~. 
\la{valUCkap}
\end{equation}

{}From (\ref{GatMG}), it is easy to derive 
$G^{d\hat{d}}$ and $G^{e^{-}e^+}$ at the scale $M_Z$. The result is
\beq
G^{d\hat{d}}(M_Z)={\rm Diag }\l \ka^{d\hat{d}}_1\si^4 ,~
-\ka^{d\hat{d}}_2\si^2 ,~ 1\r \cdot \lam_b(M_Z)~,
\la{LDatMZ}
\end{equation}
\beq
G^{e^{-}e^+}(M_Z)={\rm Diag }\l \ka^{e^{-}e^+}_1\si^4 ,~
-\ka^{e^{-}e^+}_2\si^2 ,~ 1\r \cdot \lam_{\tau }(M_Z)~,
\la{LEatMZ}
\end{equation}
with
\beqn
&\ka^{d\hat{d}}_1=\rho_b(M_Z)\rho_t^{-1}(M_Z)
\rho_{\ov{\eta }_A}(M_A)~,\non \\
&\ka^{d\hat{d}}_2=\rho_b(M_Z)\rho_t^{-1}(M_Z)
\rho_{\ov{\eta }_A}(M_A)\rho^{-1}_{\hat{\eta }_A}(M_A)~,
\la{kaD12}
\eeqn
\beqn
&\ka^{e^{-}e^+}_1=\fr{1}{\ka_u}P_{L_3}P_{AL}^3\rho^{3/2}_{\tau
}(M_Z)~,\non \\
&\ka^{e^{-}e^+}_2=\fr{1}{\ka_c}P_{L_3}\rho^{3/2}_{\tau }(M_Z)~.
\la{kaE12}
\eeqn
The numerical values of these factors are
\beqn
&\ka^{d\hat{d}}_1=1.023~,~~~\ka^{d\hat{d}}_2=0.984~,\non \\
&\ka^{e^{-}e^+}_1=0.804~,~~~\ka^{e^{-}e^+}_2=0.848~.
\la{valkaDE}
\eeqn

\subsection*{A3.~Renormalization of $A^Q$, $A^L$ matrix elements}

At the scale $M_{\rm GUT}$ the antisymmetric matrix $A$, which describes
the
fermion couplings with the components of the Higgs field $H_A$, 
is postulated to have the form
\begin{equation}
\begin{array}{cc}
 & {\begin{array}{cc}
 \hspace{-0.1cm}\hspace{2.2cm}& \hspace{-0.5cm}
\end{array}}\\ \vspace{2mm}
A(M_{\rm GUT})= \hs{-0.2cm}
\begin{array}{c}
 \\  
\end{array}\!\!\!\!\!\! &{\left(\begin{array}{ccc}
0~,&{\rm i}\si ~, &-{\rm i}\si
\\
-{\rm i}\si ~, &0~, &\fr{{\rm i}}{\sq{2}}
\\
{\rm i}\si ~,&-\fr{{\rm i}}{\sq{2}}~, &0
\end{array} \right)\lam_A\sq{2}\! }~.
\end{array}  
\label{f1Aonce}
\end{equation}
Between the scales $M_I$ and $M_{\rm GUT}$, instead of one
matrix $A$ we are dealing with the
two matrices $A^Q$ and $A^L=\bar A^L$ in the Yukawa
interaction of (\ref{yuk333}). Due to RG effects, they will differ from
the original matrix $A(M_{\rm GUT})$. $A^Q$ and $A^L$ remain antisymmetric
above $M_I$ and have the forms:
\bwi
\begin{equation}
\begin{array}{cc}
 & {\begin{array}{cc}
 \hspace{-0.1cm}\hspace{2.2cm}& \hspace{-0.5cm}
\end{array}}\\ \vspace{2mm}
A^Q= \hs{-0.1cm}{\rm i}\hs{-0.3cm}
\begin{array}{c}
 \\  
\end{array}\!\!\!\!\!\! &{\left(\begin{array}{ccc}
0~,&\ka^Q_a\si ~, &-\ka^Q_b\si
\\
-\ka^Q_a\si ~, &0~, &\fr{1}{\sq{2}}
\\
\ka^Q_b\si ~,&-\fr{1}{\sq{2}} ~, &0
\end{array} \right)\lam_A\sq{2}\! }~,~~~
\end{array}  
\begin{array}{cc}
 & {\begin{array}{cc}
 \hspace{-0.1cm}\hspace{2.2cm}& \hspace{-0.5cm}
\end{array}}\\ \vspace{2mm}
A^L= \hs{-0.1cm}{\rm i}\hs{-0.3cm}
\begin{array}{c}
 \\  
\end{array}\!\!\!\!\!\! &{\left(\begin{array}{ccc}
0~,&\ka^L_a\si ~, &-\ka^L_b\si
\\
-\ka^L_a\si ~, &0~, &\fr{1}{\sq{2}}
\\
\ka^L_b\si ~,&-\fr{1}{\sq{2}}~, &0
\end{array} \right)\lam_A^L\sq{2}\! }~.
\end{array}  
\label{matrAQLMIMG}
\end{equation} 
\ewi
Here $\ka^{Q,L}$ and $\lam_A, \lam_A^L$ are scale dependent
renormalization factors, which can
be determined through the RG equations of the $A^Q$ and $A^L$ matrices:
\beqn
&16\pi^2(A^Q)'=3A^QA^{Q\da }A^Q+\fr{3}{2}\lam_Q\lam_Q^{\da }A^Q+
\fr{3}{2}A^Q\lam_Q\lam_Q^{\da }\non \\
&+A^Q\l 3{\rm Tr}(A^QA^{Q\da })-8g_{L,R}^2-
8g_C^2\r ~,
\la{RGAQ}
\eeqn
\beqn
&16\pi^2(A^L)'=\fr{1}{2}\lam_L\lam_L^{\da }A^L+
\fr{1}{2}A^L\lam_L\lam_L^{\da }-A^L16g_{L,R}^2 +\non \\
&\te (\mu \hs{-0.1cm}-\hs{-0.1cm}M_6)A^L\l 3A^{L\da 
}A^L\hs{-0.1cm}+\hs{-0.1cm}\fr{1}{2}{\rm Tr}(A^LA^{L\da })
\hs{-0.1cm}-\hs{-0.1cm}\fr{14}{3}g_{L,R}^2\r ~.
\la{RGAL}
\eeqn 
At the scale $M_{\rm GUT}$, the boundary conditions are
\beq
A^Q(M_{\rm GUT})=A^L(M_{\rm GUT})~,~~~~{\rm and}~~~~\ka^Q_{a,b}=\ka^L_{a,b}=1~.
\la{boundAMG}
\end{equation} 
{}From (\ref{RGAQ}) and (\ref{RGAL}) follow the RG equations 
(\ref{RGAMI}) and (\ref{RGSMI}) for
$\lam_A$
and $\lam_A^L$, which we solved numerically. It is easy to observe,
that the ratios $A^Q_{13}/A^Q_{23}$ and $A^L_{13}/A^Q_{23}$ do not run,
while the other ratios are scale dependent. For the
factors $\ka^{Q,L}$ we have at $\mu =M_I$
\beq
\ka^Q_a=P_{Q_3}^{3/2}~,~~~\ka^L_a=P_{L_3}^{1/2}~,~~~
\ka^Q_b=\ka^L_b=1~,
\la{kaQLMI}
\end{equation}
where $P_{Q_3}$, $P_{L_3}$ are given in (\ref{Pfactors}),
(\ref{numPfact}).

Below the scale $M_I$, instead of one $A^Q$ matrix there are matrices
which represent the couplings of colored fermions with Higgs
doublets and singlet. Namely:
\beqn
&Q_LA^QQ_R~H_A\to A^{d\hat{d}}q\hat{d}~(H_A)_2^{1,2}+
A^{D\hat{d}}D\hat{d}~(H_A)^3_2\non \\
&+A^{d\hat{D}}q\hat{D}~(H_A)_3^{1,2}~.
\la{Adec}
\eeqn 
At the scale $M_I$ all matrices unify 
$A^{d\hat{d}}(M_I)=A^{D\hat{d}}(M_I)=
A^{d\hat{D}}(M_I)=A^Q(M_I)$. There they are
precisely antisymmetric and their matrix elements match with the
appropriate $\ka^Q_{a, b}$ factors of (\ref{matrAQLMIMG}).
RG study allow to calculate these matrices at the scale needed. 
Our numerical
analysis gives the following results for the $A^Q$-matrices  at the scale
$\mu =M_Z$ (where contact with experimental data can be performed):
\begin{equation}
\begin{array}{cc}
 & {\begin{array}{cc}
 \hspace{-0.1cm}\hspace{2.2cm}& \hspace{-0.5cm}
\end{array}}\\ \vspace{2mm}
A^{d\hat{d}}(M_Z)\simeq \hs{-0.1cm}{\rm i}\hs{-0.3cm}
\begin{array}{c}
 \\  
\end{array}\!\!\!\!\!\! &{\left(\begin{array}{ccc}
0~,&1.16\si ~, &-\si
\\
-1.16\si ~, &0~, &\fr{1}{\sq{2}}
\\
0.99\si ~,&-\fr{0.99}{\sq{2}} ~, &0
\end{array} \right)\ov{\lam }_A(M_Z)\sq{2}\! }~,
\end{array}  
\la{matrAdec1}
\end{equation}
\beq
\begin{array}{cc}
 & {\begin{array}{cc}
 \hspace{-0.1cm}\hspace{2.2cm}& \hspace{-0.5cm}
\end{array}}\\ \vspace{2mm}
A^{D\hat{d}}(M_Z)\simeq \hs{-0.1cm}{\rm i}\hs{-0.3cm}
\begin{array}{c}
 \\  
\end{array}\!\!\!\!\!\! &{\left(\begin{array}{ccc}
0~,&1.39\si ~, &-1.2\si
\\
-1.16\si ~, &0~, &\fr{1}{\sq{2}}
\\
0.97\si ~,&-\fr{0.97}{\sq{2}}~, &0
\end{array} \right)\lam_A(M_Z) \sq{2}\! }~,
\end{array}  
\label{matrAdec2}
\end{equation} 
\beq
\begin{array}{cc}
 & {\begin{array}{cc}
 \hspace{-0.1cm}\hspace{2.2cm}& \hspace{-0.5cm}
\end{array}}\\ \vspace{2mm}
A^{d\hat{D}}(M_Z)\simeq \hs{-0.1cm}{\rm i}\hs{-0.3cm} 
\begin{array}{c}
 \\  
\end{array}\!\!\!\!\!\! &{\left(\begin{array}{ccc}
0~,&1.19\si ~, &-\si
\\
-1.42\si ~, &0~, &\fr{1}{\sq{2}}
\\
1.21\si ~,&-\fr{1.02}{\sq{2}}~, &0
\end{array} \right)\lam_A^0(M_Z)\sq{2}\! }~.
\end{array}  
\la{matrAdec3}
\end{equation}
The couplings are
\beqn
&\ov{\lam }_A(M_Z)=1.262~,~~~~\lam_A(M_Z)=1.131~,\non \\
&\lam_A^0(M_Z)=1.220~.
\la{AQ23MZ}
\eeqn

A similar analysis can be performed to obtain the generation
mixing matrices $A^L$ for leptons. 
Below $M_I$ we are dealing with three types of matrices
$A^{e^{-}e^{+}}$, $A^{E^{-}e^{+}}$ and $A^{e^{-}E^{+}}$. 
These coupling matrices are relevant for the charged lepton
sector. At the scale $\mu =M_Z$ we find
\begin{equation}
\begin{array}{cc}
 & {\begin{array}{cc}
 \hspace{-0.1cm}\hspace{2.2cm}& \hspace{-0.5cm}
\end{array}}\\ \vspace{2mm}
A^{e^{-}e^{+}}(M_Z)\simeq  \hs{-0.1cm}{\rm i}\hs{-0.3cm}
\begin{array}{c}
 \\  
\end{array}\!\!\!\!\!\! &{\left(\begin{array}{ccc}
0~,&1.11\si ~, &-\si
\\
-1.11\si ~, &0~, &\fr{1}{\sq{2}}
\\
1.04\si ~,&-\fr{1.04}{\sq{2}} ~, &0
\end{array} \right)\ov{\lam }_A^L(M_Z)\sq{2}\! }~,
\end{array}  
\la{matrALdec1}
\end{equation}
\beq
\begin{array}{cc}
 & {\begin{array}{cc}
 \hspace{-0.1cm}\hspace{2.2cm}& \hspace{-0.5cm}
\end{array}}\\ \vspace{2mm}
A^{E^{-}e^{+}}(M_Z)\simeq \hs{-0.1cm}{\rm i}\hs{-0.3cm}
\begin{array}{c}
 \\  
\end{array}\!\!\!\!\!\! &{\left(\begin{array}{ccc}
0~,&1.14\si ~, &-1.03\si
\\
-1.11\si ~, &0~, &\fr{1}{\sq{2}}
\\
1.06\si ~,&-\fr{1.06}{\sq{2}}~, &0
\end{array} \right)\hs{-0.1cm}\lam_A^L(M_Z)\sq{2}\! }~,
\end{array}  
\label{matrALdec2} 
\end{equation} 
\beq
\begin{array}{cc}
 & {\begin{array}{cc}
 \hspace{-0.1cm}\hspace{2.2cm}& \hspace{-0.5cm}
\end{array}}\\ \vspace{2mm}
A^{e^{-}E^{+}}(M_Z)\simeq \hs{-0.1cm}{\rm i}\hs{-0.3cm} 
\begin{array}{c}
 \\  
\end{array}\!\!\!\!\!\! &{\left(\begin{array}{ccc}
0~,&1.07\si ~, &-1.05\si
\\
-1.12\si ~, &0~, &\fr{1}{\sq{2}}
\\
1.06\si ~,&-\fr{1.01}{\sq{2}}~, &0
\end{array} \right)\hs{-0.1cm}\lam_A^{0L}(M_Z)\sq{2}\! }~.
\end{array}  
\la{matrALdec3}
\end{equation}
The coupling factors are
\beqn
&\ov{\lam }_A^L(M_Z)=0.957~,~~~~
\lam_A^L(M_Z)=0.923~,\non \\
&\lam_A^{0L}(M_Z)=1.068~.
\la{23ALMZ}
\eeqn 

The picture which shows the running of all $(2,3)$ and $(3,2)$ elements
of the matrices $A^Q$ and $A^L$
and  their final unification is presented in figure 3, the 'desert
spider'.

\subsection*{A4.~Neutrino mass matrix renormalization}

Here we will perform the renormalization analysis for the neutrino sector.

Dimension five operators, responsible for neutrino masses, are generated
by integrating out the 'right handed'
states $(L^3_2)_{\al }=\hat{\nu }_{\al }$. The $\hat{\nu }_{\al }$
masses are determined by the matrix $F^{\{2,2\}}S$. 
The  matrix $S$ describes the generation dependent Majorana couplings of
these states to the symmetric sextet component of the Higgs field
$H_S$. In our model $S$ is postulated to be the bilinear matrix product in
generation space (\ref{bilin}).
We take this form to be valid at $M_I$ with $G\to G_L$ and $A\to A^L$.
With the appropriate scaling
factors, at $\mu =M_I$ the matrix $S$ has the form:
\begin{equation}
\begin{array}{cc}
 & {\begin{array}{cc}
 \hspace{-0.1cm}\hspace{2.2cm}& \hspace{-0.5cm}
\end{array}}\\ \vspace{2mm}
S/\lam_S\simeq \hs{-0.2cm}
\begin{array}{c}
 \\  
\end{array}\!\!\!\!\!\! &{\left(\begin{array}{ccc}
(\ka^L_1)^2\si^8~,&-\ka^L_2\ka^L_a\si^6x ~, &-\si^4x
\\
-\ka^L_2\ka^L_a\si^6x ~, &(\ka^L_2)^2\si^4~, &\fr{\si^3}{\sq{2}}x
\\
-\si^4x ~,&\fr{\si^3}{\sq{2}}x~, &1
\end{array} \right)\! }~.
\end{array}  
\label{SformMI}
\end{equation}   
Again, as in (\ref{Sform}), only the leading terms in $\si $ are 
exhibited here.
The renormalization coefficients $\ka $ are shown in (\ref{ka12a}).
At the scale $M_I$, the mass matrix for the states $\hat{\nu }_{\al }$ is 
\beq
M_{\hat{\nu }\hat{\nu }}^{\al \bt }(M_I)=F^{\{2,2\}}S^{\al \bt }~.
\la{nuRmas}
\end{equation}
The eigenvalues are 
$\mu_1\simeq \si^8\l (\ka^L_1)^2-(\ka^L_a)^2x^2-x^2+
\fr{\ka_a^L}{\ka^L_2}\sq{2}x^3\si \r \lam_SF^{\{2,2\}}$,
$\mu_2\simeq \si^4(\ka^L_2)^2\lam_SF^{\{2,2\}}$ and
$\lam_SF^{\{2,2\}}$. As we can
see, these scales are separated by large distances. Because of this fact
strong renormalization effects occur. They are the cause of the difference 
between the neutrino mass matrices (\ref{Mnubare}) and
(\ref{Mnumatr}). 
The decoupling of the three 
$\hat{\nu }_{\al }$ states occurs step by step. Thus the renormalization
has to be performed separately in each energy interval 
\cite{lindner}, \cite{ratz}.
By generalizing the results of ref. \cite{ratz} we present model 
independent formula for the running
of the neutrino mass matrix
$m_{\nu }$ and then apply them to our model.

The couplings in the neutrino sector involve Dirac and Majorana mass
terms. One can choose a basis in which the mass matrix
for the heavy neutrinos is
diagonal. Thus, without loss of generality, one can write the coupling
terms in the form
\beq
\nu_{\al }(\lam^{\rm Dirac})^{\al \bt}\hat{\nu }_{\bt }+
\fr{1}{2}\mu_{\al }\hat{\nu }_{\al }\hat{\nu }_{\al }~.
\la{DirMaj}
\end{equation}
The matrix $\lam^{\rm Dirac} $ is related to the original Dirac matrix 
$\lam_0^{\rm Dirac}=G^{\nu \hat{\nu }}e_1^1=G_Le_1^1$
(for $\mu \stackrel{>}{_-}M_I$) via
\beq
\lam^{\rm Dirac} =\lam_0^{\rm Dirac}U_S^T~.
\la{relDir}
\end{equation}
Here the unitary matrix $U_S$ diagonalizes the matrix 
$M_{\hat{\nu }\hat{\nu }}$:
\beq
\l {U_SM_{\hat{\nu }\hat{\nu }}U_S^T}\r^{\al \bt}=
\de^{\al \bt}\mu_{\al }~.
\la{diagMaj}
\end{equation}
By integrating out the state $\hat{\nu }_{\al }$, the light neutrino mass
matrix gets a contribution at the scale $\mu_{\al }$. 
Without renormalization effects, $m_{\nu }$ would have the form
\beq
m_{\nu }^{\al \bt}=-(Y_1+Y_2+Y_3)^{\al \bt}~,
\la{wouldMnu}
\end{equation}
where
\beqn
&Y_1^{\al \bt}=\fr{1}{\mu_1}(\lam^{\rm Dirac})^{\al 1}
(\lam^{\rm Dirac})^{\bt 1}~,\non \\
&Y_2^{\al \bt}=\fr{1}{\mu_2}(\lam^{\rm Dirac})^{\al 2}
(\lam^{\rm Dirac})^{\bt 2}~,\non \\
&Y_3^{\al \bt}=\fr{1}{\mu_3}(\lam^{\rm Dirac})^{\al 3}
(\lam^{\rm Dirac})^{\bt 3}~.
\la{defYi}
\eeqn
The division of the mass matrix in three parts is convenient
in order to see the
contributions coming from each integrated state $\hat{\nu }_{\al }$.
Each $d=5$ operator ($Y_i$), generated on the scale $\mu_i$, runs from
this scale down to $M_Z$ according to the RG equations
\beq
4\pi \fr{d}{dt}Y_i^{\al \bt}=Y_i^{\al \bt}(6\eta_t-3\al_2)~,
~~~~{\rm with}~~~(\al ,\bt )\neq (3, 3)~, 
\la{RGYij}
\end{equation}
\beq
4\pi \fr{d}{dt}Y_i^{3\al }=Y_i^{3\al }(6\eta_t+
\fr{1}{2}\eta_{\tau }-3\al_2)~,
~~~~{\rm with}~~~\al \neq 3~,
\la{RGY3i}
\end{equation}
\beq
4\pi \fr{d}{dt}Y_i^{33}=Y_i^{33}(6\eta_t+\eta_{\tau }-3\al_2)~.
\la{RGY33}
\end{equation}
These equations have the solutions
\beq
Y_i^{\al \bt }(\mu')=Y_i^{\al \bt }(\mu )r_g(\mu')r_g^{-1}(\mu )~ ,
\la{solYij}
\end{equation}
\beq
Y_i^{3\al }(\mu')=Y_i^{3\al}(\mu )r_g(\mu')r_{\tau }(\mu')
r_g^{-1}(\mu )r_{\tau }^{-1}(\mu )~ .
\la{solY3i}
\end{equation}
\beq
Y_i^{33}(\mu')=Y_i^{33}(\mu )r_g(\mu')r_{\tau }^2(\mu')
r_g^{-1}(\mu )r_{\tau }^{-2}(\mu )~,
\la{solY33}
\end{equation}
where
\beq
r_g(\mu )=\rho_t^{-6}(\mu )\rho_{\al_2}^3(\mu )~,~~~~~
r_{\tau }(\mu )=\rho_{\tau }^{-1/2}(\mu )~.
\la{rgrtau}
\end{equation}
Before the emergence of the $d=5$ operators, only the Dirac couplings
run. They obey the equations \cite{ft3}
\beq
4\pi \fr{d}{dt}(\lam^{\rm Dirac})^{\al \bt }=0 ~,
~~~~~~~~~(\al, \bt )\neq (3, 3)~,
\la{RGdiracij}
\end{equation}
\beq
4\pi \fr{d}{dt}(\lam^{\rm Dirac})^{3\al }=\lam^{3 \al }
\fr{1}{2}\eta_{\tau }~,  
~~~~~~~~\al \neq 3~.
\la{RGdirac3i}
\end{equation}
According to these equations, we have for $\mu <M_I$
\beqn
&(\lam^{\rm Dirac})^{\al \bt }(\mu )=
(\lam^{\rm Dirac})^{\al \bt }(M_I)~,\non \\
&(\lam^{\rm Dirac})^{3 \al }(\mu )=(\lam^{\rm Dirac})^{3 \al }(M_I)
r_{\tau }(\mu )~.
\la{solN}
\eeqn 
With all these results, one can write down the light neutrino mass
matrix at the scale $\mu =M_Z$
\beq
m_{\nu }^{\al \bt }(M_Z)=-{\cal Y}^{\al \bt }r_g(M_Z)~,
\la{MnuMZ}
\end{equation}
where
\bwi
\beqn
&{\cal Y}^{\al \bt }=\sum_{i,j=1,2}\de^{\al i}\de^{\bt j}
\l Y_3^{ij}+r_g^{-1}(\mu_2)Y_2^{ij}+r_g^{-1}(\mu_1)Y_1^{ij}\r +\non \\
&\sum_{i=1,2}\l \de^{\al 3}\de^{\bt i}+\de^{\bt 3}\de^{\al i}\r
r_{\tau }(M_Z)
\l Y_3^{i3}+r_g^{-1}(\mu_2)Y_2^{i3}+r_g^{-1}(\mu_1)Y_1^{i3}\r \non \\
&\de^{\al 3}\de^{\bt 3}r_{\tau }^2(M_Z)
\l Y_3^{33}+r_g^{-1}(\mu_2)Y_2^{33}+r_g^{-1}(\mu_1)Y_1^{33}\r ~.
\la{calY}
\eeqn
\ewi 
The quantities
$Y_i$ are given in (\ref{defYi}) and the renormalization factors
in (\ref{rgrtau}).

We now  apply this result to our model. 
The matrix $(\lam^{\rm Dirac}) $ is built according to
(\ref{relDir}) taking
\beq
\lam_0^{\rm Dirac}(M_I)={\rm Diag}\l \ka^L_1\si^4 ,~-\ka^L_2\si^2 ,~1\r 
\lam_{\tau }(M_I)e^1_1(M_I)~.
\la{nuDirac}
\end{equation}
The value of $e_1^1$ at the scale $M_I$ can be calculated from the RG
equation
\beq
4\pi \fr{d}{dt}e_1^1=e_1^1 
\l -3\eta_t+\fr{9}{20}\al_1+\fr{9}{4}\al_2\r ~,
\la{RGe11}
\end{equation}
knowing its value at the scale $\mu =M_Z$:
\beq	
e_1^1(M_I)=e_1^1(M_Z)
\rho_t^{-3}(M_Z)\rho_{\al_1}^{9/20}(M_Z)\rho_{\al_2}^{9/4}(M_Z)~.
\la{1sole11}
\end{equation}
In our model, it turns out that $e_1^1(M_I)\simeq 124.52$~GeV.

The values of the factors appearing in (\ref{SformMI}) are
\beq
\kappa_1^L\simeq 0.707~,~~~~
\kappa_2^L\simeq 0.746~,~~~~ 
\kappa_a^L\simeq 1.017~.
\la{ka12a}
\end{equation}
After constructing the matrix $\lam^{\rm Dirac} $, one can
calculate all elements of
$Y_i$ according to (\ref{defYi}). The numerical values of the
renormalization factors, appearing in (\ref{MnuMZ}), (\ref{calY})
are
\beqn
&r_g(M_Z)\simeq 0.58~,~~~
r_g(\mu_1)\simeq 0.75~,\non \\
&r_g(\mu_2)\simeq 0.87~,~~~
r_{\tau }(M_Z)\simeq 0.96~.
\la{rInMnu}
\eeqn 

The diagonalization of the neutrino mass matrix obtained in this way 
allows to
calculate the neutrino mixing and the ratio 
$R=\De m^2_{\rm sol}/\De m^2_{\rm atm}$ from the single parameter $x$.
To obtain values for $R$ and the mixing angles, which lie within 
experimental bounds, we have to use $x\simeq 2.8$
which is bit smaller than the value $x=3.5$ found without renormalization
corrections \cite{ft4}. 
The mixing is now no more bimaximal, but still bilarge.
The results are discussed in the text (section \ref{sec:nu}).

\newpage 
\bibliography{apssamp}

\begin{thebibliography}{99}
\bibitem{ft1}
We took the quark fields to be color
antitriplets, the antiquark fields color triplets. In this way the
cyclic symmetry of $E_6$ is obvious \cite{eric}.
\bibitem{ft3}
The running of $\lam^{33}$ is not
relevant because it involves terms connected with integrating out the 
$\hat{\nu }_3$ state. We assume that all Yukawa couplings related to 
Dirac masses $\lam^{\al \bt }$ [$(\al ,\bt )\neq (3,3)$] are small.
Also, it is assumed that Yukawa couplings
involving $\hat{\nu }\hat{\nu }$ terms are small. Therefore
the scale factors 
$\mu_{\al }$ do not run in the energy interval we are dealing with
(these assumptions are valid in the model considered).
\bibitem{ft4}
The scaling factors in (\ref{rInMnu}) have a mild $x$
dependence. The values presented correspond to $x\simeq 2.8$.
\end{thebibliography}

\bibliographystyle{unsrt}

\end{document}